\documentclass[11pt,a4paper]{article}
\usepackage{boondox-cal}
\usepackage{float}
\usepackage{diagbox}
\usepackage{jheppub}
\usepackage{graphicx}
\usepackage{amsmath}
\usepackage{amssymb}
\usepackage{amsthm}
\usepackage{mathtools}
\usepackage{mathrsfs}
\input epsf
\usepackage{epsfig}
\usepackage{tikz}
\usepackage{subfigure,tikz}
\usetikzlibrary{backgrounds,automata}
\usepackage{graphics}
\usepackage{subfigure}
\usepackage[active]{srcltx}
\usepackage[T1]{fontenc} 
\usepackage{graphicx}
\usepackage{natbib}
\usepackage{hyperref}
\newcommand{\Autoref}[1]{Section~\ref{#1}}
\usepackage[title]{appendix}

\newcommand{\bea}{\begin{eqnarray}}
	\newcommand{\eea}{\end{eqnarray}}
\newcommand{\bean}{\begin{eqnarray*}}
	\newcommand{\eean}{\end{eqnarray*}}

\allowdisplaybreaks

\def\Label#1{\label{#1}%
	\smash{\hbox to0pt{\raise1ex\hbox{\tiny[#1]}\hss}}}

\newcommand{\parall}[2]{{#1}\ /\kern -0.8em / \  {#2}}

\title{\boldmath An Explicit Expression of Generating Function for One-Loop Tensor Reduction}
\preprint{USTC-ICTS/PCFT-23-32}

\author[a,b]{Chang Hu*,}

\affiliation[a]{School of Fundamental Physics and Mathematical Sciences, Hangzhou Institute for Advanced Study, UCAS, Hangzhou 310024, China}

\affiliation[b]{University of Chinese Academy of Sciences, Beijing 100049, China}

\author[d,e]{Tingfei Li,}

\author[a,b,c]{Jiyuan Shen*,}

\affiliation[c]{Institute of Theoretical Physics, Chinese Academy of Sciences, Beijing 100190, China}

 \affiliation[d]{Zhejiang Institute of Modern Physics, School of Physics, Zhejiang University, \\Hangzhou, Zhejiang 310058, China }
\affiliation[e]{Joint Center for Quanta-to-Cosmos Physics, Zhejiang University,
\\Hangzhou, Zhejiang 310058, China}

\author[f,g]{Yongqun Xu}

\affiliation[f]{Interdisciplinary Center for Theoretical Study, University of Science and Technology of
China, Hefei, Anhui 230026, China}

\affiliation[g]{Department of Physics, South China University of Technology, Guangzhou 510641, China}

\emailAdd{isiahalbert@126.com} \emailAdd{tfli@zju.edu.cn} \emailAdd{isaiahyshen@gmail.com} \emailAdd{yongqunxu@mail.ustc.edu.cn}

\abstract{ This work introduces an explicit expression for the generating function for the reduction of an $n$-gon to an $(n-k)$-gon. A novel recursive relation of generating function is formulated based on Feynman Parametrization in projective space, involving a single ordinary differential equation.  The explicit formulation of generating functions provides crucial insights into the complex analytic structure inherent in loop amplitudes.

}

\renewcommand{\thefootnote}{\fnsymbol{footnote}}
\footnotetext{*Corresponding author}

\usepackage{amsfonts}
\begin{document}
\maketitle

\renewcommand{\thefootnote}{\arabic{footnote}}
\setcounter{footnote}{0} 
	\flushbottom
	
\section{Background and Motivation}

Scattering amplitudes are fundamental concepts in modern quantum field theory. They serve as not only a source of precise predictions and explanations for experiments conducted at the LHC \cite{NLOMultilegWorkingGroup:2008bxd} but also make a significant contribution to the analysis of the analytical structure within perturbative quantum field theory. Moreover, they play a pivotal role in our exploration of potential new phenomena in physics. Amplitudes, which result from perturbative expansions, can be classified into two primary categories: tree diagrams and loop diagrams, based on their respective orders.

The study of tree diagrams has a rich history, and various discoveries, such as the Bern-Carrasco-Johansson (BCJ) color-kinematic duality\cite{Bern:2008qj,Chiodaroli:2014xia,Johansson:2015oia,Johansson:2019dnu}, the Kawai-Lewellen-Tye (KLT) relation\cite{Kawai:1985xq}, and on-shell techniques like the Britto-Cachazo-Feng-Witten (BCFW) recursion relations\cite{Britto:2004ap,Britto:2005fq} have significantly deepened our comprehension of field theory. More recently, there have been notable developments like the Cachazo-He-Yuan (CHY) formalism\cite{Cachazo:2013gna,Cachazo:2013hca,Cachazo:2013iea,Cachazo:2014nsa,Cachazo:2014xea}, which strives to reframe field theory from twistor string\cite{Witten:2003nn}.

At the level of loop diagrams, in the context of constructing Feynman integral expressions, one of the most significant methods over the past few decades has been the utilization of on-shell techniques\cite{Henn,Elvang,Bern:1994zx,Bern:1994cg,Cachazo:2004kj,Britto:2004nc,Britto:2004ap,Britto:2005fq,Ellis:2011cr}. On-shell methods are grounded in a fundamental physical concept: when a propagator becomes on-shell, a physical process factorizes into the product of two physical processes. Building upon this concept, the unitarity cut method \cite{Bern:1994zx,Bern:1994cg,Ellis:2011cr,Britto:2004nc,Britto:2005ha} has been developed, allowing the calculation of a one-loop diagram to be reduced to a product of several tree-level diagrams. This approach significantly simplifies the computations compared to the Feynman rules. In addition to this, within the CHY framework, there have been relevant developments in constructing Feynman integrands for loop diagrams. These efforts primarily involve taking the forward limit of two external legs in the $(n+2)$-point tree-level amplitudes. This approach establishes a relationship between $n$-point one-loop diagrams and $(n+2)$-point tree-level amplitudes\cite{Mason:2013sva,Adamo:2013tsa,Adamo:2015hoa,Casali:2014hfa,Geyer:2015bja,Geyer:2015jch,Geyer:2017ela,He:2015yua,Cachazo:2015aol,Feng:2019xiq}.

After providing the loop Feynman integrand, the usual approach for computing this integral involves expressing it as a linear combination of a set of basis integrals, with the coefficients being rational functions of external momenta, masses, and spacetime dimension\cite{Weinzierl:2022eaz}. These basis integrals, referred to as master integrals, can be computed once and for all. The process of reduction can be divided into two categories: integrand-level reduction and integral-level reduction. The integrand-level reduction can indeed be viewed as an algebraic factorization of the loop integrand. It can be systematically resolved using computational algebraic geometry\cite{Ossola:2006us,Mastrolia:2011pr,Badger:2012dp,Zhang:2012ce,Larsen:2015ped,Zhang:2016kfo}. As for the integral-level reduction, the initial proposal was the notable Passarino-Veltman reduction (PV-reduction) method\cite{Passarino:1978jh}. By observing the possible tensor structures and employing the method of undetermined coefficients, we can recursively determine the coefficients for each tensor structure. Other proposals include the Integration-by-Parts (IBP) method\cite{Schabinger:2011dz,Chetyrkin:1981qh,Tkachov:1981wb,Laporta:2000dsw,vonManteuffel:2012np,vonManteuffel:2014ixa,Maierhofer:2017gsa,Smirnov:2019qkx}, the unitarity cut method\cite{Bern:1994zx,Bern:1994cg,Britto:2004nc,Britto:2005ha,Britto:2006sj,Anastasiou:2006jv,Anastasiou:2006gt,Britto:2006fc,Britto:2007tt,Britto:2010um}, and the Intersection number method\cite{Mastrolia:2018uzb,Frellesvig:2019uqt,Mizera:2019vvs,Frellesvig:2020qot,Caron-Huot:2021xqj,Caron-Huot:2021iev,Chen:2023kgw}. Despite numerous advancements in integral-level reduction, the growing complexity of computations warrants further improvements. There are some mature software packages available for calculating higher-loop\cite{Wu:2023upw, Peraro:2019svx, Klappert:2019emp, Klappert:2020aqs, Belitsky:2023qho, Liu:2018dmc, Guan:2019bcx, www:Blade, Smirnov:2020quc, Usovitsch:2020jrk, Anastasiou:2004vj, Smirnov:2008iw, Smirnov:2013dia, Smirnov:2014hma, Smirnov:2019qkx, Maierhofer:2017gsa, Maierhofer:2018gpa, Klappert:2020nbg, Lee:2012cn, Lee:2013mka, Studerus:2009ye, Smirnov:2008py, Smirnov:2009pb, Smirnov:2013eza, Smirnov:2015mct, Smirnov:2021rhf}. 

This article will primarily focus on the tensor reduction problem at one-loop level. This problem has been under investigation for an extended period; however, a recent breakthrough reveals that the introduction of a virtual auxiliary vector $R$, can streamline the conventional Passarino-Veltman reduction process\cite{Feng:2021enk,Hu:2021nia,Feng:2022uqp,Feng:2022iuc,Feng:2022rfz}. When an auxiliary vector $R$ is incorporated into the IBP method, the improvement of efficiency of reduction has been shown in\cite{Chen:2022jux,Chen:2022lue,Chen:2022lzr}.

Another notable advancement is the introduction of the "generating function" \cite{Ablinger:2014yaa,Kosower:2018obg,Behring:2023rlq,Feng:2022hyg,Guan:2023avw}. In fact, the concept of a generating function is well-established in both physics and mathematics. The application of generating functions in higher-order QFT calculations was introduced earlier in \cite{Ablinger:2014yaa} and has been utilized up to the level of $3$- and $4$-loop calculations. In \cite{Kosower:2018obg}, Kosower employed integration-by-parts (IBP) generating vectors to derive appropriate IBP reduction relations in certain special $2$-loop case. More recently, these functions have also been discussed in \cite{Behring:2023rlq}, detailing how to directly proceed from these representations to the physical result.

In the reduction process, the reduction coefficients for high tensor rank still present a considerable challenge. However, when we sum up the reduction coefficients of different tensor ranks, we might arrive at a simpler solution. For instance, in our method, the numerator of the integral is  $(2l\cdot R)^{r}$
 with rank $r$. We can sum them up in two typical ways as shown in the equations below:

\begin{equation}
    \psi_1(\mathcal{t})=\sum_{r=0}^\infty \mathcal{t}^r\cdot (2l\cdot R)^r=\frac{1}{1-\mathcal{t}(2l\cdot R)},\ \ \psi_2(\mathcal{t})=\sum_{r=0}^\infty \frac{(2l\cdot R)^r\cdot\mathcal{t}^r}{r!}=e^{\mathcal{t}\cdot(2l\cdot R)}.
\end{equation}

In recent research\cite{Feng:2022hyg,Guan:2023avw}, Bo Feng proposed a recursive method to calculate the generating function, establishing several partial differential equations based on the auxiliary $\textbf{R}$ method. Subsequent work\cite{Guan:2023avw} has not only improved upon this at the one-loop level but also extended it to the $2$-loop level by setting up and solving differential equations of these generating functions, utilizing the Integration-by-Parts (IBP) method. Both pieces of work underscore the considerable potential of generating functions. However, in both studies, authors provide only an iterative approach to compute the generating functions, and it remains strenuous to directly write generating functions explicitly, even at the one-loop level. Fortunately, we discovered a new recursive relation of generating functions based on investigations into Feynman parametrization in the projective space\cite{Feng:2022rwj,Li:2022cbx,Feng:2022rfz}. This new relation consists of a single ordinary differential equation on $\mathcal{t}$ instead of a complex set of partial differential equations, allowing us to directly write an \textbf{explicit expression} of the generating function for the reduction of $n$-gon to $(n-k)$-gon for \textbf{universal} $k$ without recursion.

The organization of this paper is as follows. \Autoref{sec:prep} introduces essential notations used throughout the paper and establishes our new recursive relation. In \Autoref{sec:n to n}, we compute the generating function of $n$-gon to $n$-gon as a warm-up. \Autoref{sec:n to n-1} and \Autoref{sec:n to n-2} detail the derivation of the generating function of $n$-gon to $(n-1)$-gon and $n$-gon to $(n-2)$-gon, respectively, as two non-trivial examples. After summarizing the previous results in \Autoref{sec:n to n-k}, we provide an explicit expression of the generating function for $n$-gon to $(n-k)$-gon of universal $k$. \Autoref{sec:proof} offers an inductive proof  of our results. Finally, \Autoref{sec:con} provides a brief summary and discussion. Beside, we present the solution of  the typical differential equation in Appendix \ref{sec:int}. We also provide the  numerical verification in Appendix \ref{sec:num check}.

	\section{Preparation}
	\label{sec:prep}

	In this paper, our goal is to provide an explicit expression for the generating functions of one-loop tensor reduction. As we know, by introducing an auxiliary vector $R$, the general one loop integral in $D$ dimensions can be written as 
	
	\begin{equation}
		I^{(r)}_n=\int\frac{d^Dl}{i\pi^{D/2}}\frac{(2R\cdot l)^r}{\prod^n_{j=1}(l-q_j)^2-M_j^2}.
	\end{equation}
	In this formula, $n$
 represents the number of propagators, and $r$
 denotes the tensor rank number.

\subsection{Notations}
Let us start with the introduction of some notations that we'll be using subsequently.
	
	\begin{itemize}
		\item Some $n$-dimension vectors:
		\begin{equation}
			\begin{aligned}
				&\boldsymbol{L}:\ L_i=1,\\
				&\boldsymbol{V}:\ V_i=R\cdot q_i,\\
				&\boldsymbol{H}_b=\{0,\cdots,0,1,0,\cdots,0\},
			\end{aligned}
		\end{equation}
		where $1$ is in the $b$-th position.
		\item $Q$ is an $n\times n$ matrix defined as
		\begin{equation}
			Q_{ij}=\frac{M^2_i+M^2_j-(q_i-q_j)^2}{2}.
		\end{equation}
		
		\item The notations $(\overline{AB})$ and $(\overline{AB})_\textbf{b}$, with a label list $\textbf{b}=\{b_1,b_2,...,b_k\}$ for two vectors $A$ and $B$, are defined as follows:

		\begin{equation}
			\begin{aligned}
				(\overline{AB})=&A\cdot Q^{-1}\cdot B,\\
				(\overline{AB})_\textbf{b}=&A_{\widehat{\textbf{b}}}\cdot (Q_{\widehat{\textbf{b}}\widehat{\textbf{b}}})^{-1}\cdot B_{\widehat{\textbf{b}}}.
			\end{aligned}
		\end{equation}
		
		In this context, $A_{\widehat{\textbf{b}}}$ and $B_{\widehat{\textbf{b}}}$ are two vectors derived by removing all the $b_i$-th components from vectors $A$ and $B$. $Q_{\widehat{\textbf{b}}\widehat{\textbf{b}}}$ is a matrix obtained by eliminating all the $b_i-$th rows and $b_i-$th columns from matrix $Q$. For example, with $n=4$ and $\textbf{b}=\{2,3\}$, we have 
		
		\begin{equation}
				(\overline{VL})_{2,3}=\begin{pmatrix}
					R\cdot q_1  ,& R\cdot q_4
				\end{pmatrix}\begin{pmatrix}
					M_1^2  & \frac{M_1^2+M_4^2+(q_1-q_4)^2}{2} \\
					\frac{M_4^2+M_1^2+(q_4-q_1)^2}{2} &  M_4^2
				\end{pmatrix}^{-1}\begin{pmatrix}
				    1\\1
				\end{pmatrix}.
		\end{equation}

		\item If $\Omega$ is an analytic expression composed of $(\overline{AB})$ or $(\overline{AB})_{\textbf{b}}$, then $[\Omega]_{\textbf{a}}$  with a label set $\textbf{a}$ outside the square brackets represents the analytic expression obtained by appending subscript $\textbf{a}$ on each term of the form $(\overline{AB})$ or $(\overline{AB})_{\textbf{b}}$ present in $\Omega$. For example
		
		\begin{equation}
			\begin{aligned}
				[(\overline{VL})]_{1,2}=(\overline{VL})_{1,2},\ \ \ [(\overline{LL})_{2}]_{3}=(\overline{LL})_{2,3}.
			\end{aligned}
		\end{equation}
		If 
		\begin{equation}
			P=\frac{(\overline{H_1L})(\overline{VV})_{3}+(\overline{LL})_{2}R^2}{(\overline{VL})_{2}(H_3V)_{1}},
		\end{equation}
		
		then
		
		\begin{equation}
			[P]_{4,5}=\frac{(\overline{H_1L})_{4,5}(\overline{VV})_{3,4,5}+(\overline{LL})_{2,4,5}R^2}{(\overline{VL})_{2,4,5}(H_3V)_{1,4,5}}.
		\end{equation}

		\item $I^{(r)}_{n,\widehat{\textbf{b}}}$ represents a one-loop integral where all
propagators of $I^{(r)}_{n}$ with index $b_i\in \textbf{b}$ are removed
		
		\begin{equation}
			I^{(r)}_{n,\widehat{\textbf{b}}}=\int\frac{d^Dl}{i\pi^{D/2}}\frac{(2R\cdot l)^r}{\prod^n_{j=1,j\notin \textbf{b}}(l-q_j)^2-M_j^2}.
		\end{equation}   
	\end{itemize}

\subsection{Recursive relation}
 
After introducing the above notations, we begin to derive the recursive relations for the generating function of the reduction coefficients. As pointed out in \cite{Li:2022cbx} and \cite{Feng:2022rfz}, there exists a non-trivial recursion relation for one-loop tensor integrals for $r\geq 1$\footnote{Where $b$ represents a single label $b$.}
	
	\begin{equation}
		\begin{aligned}
			I_n^{(r)}&=\frac{2(D+2r-n-2)}{D+r-n-1}\cdot\frac{(\overline{VL})}{(\overline{LL})}\cdot I^{(r-1)}_n+\frac{4(r-1)}{D+r-n-1}\cdot \frac{R^2-(\overline{VV})}{(\overline{LL})}\cdot I_n^{(r-2)}\\
   &+\sum_{b=1}^n(X^{(b)}\cdot I^{(r-1)}_{n,\widehat{b}}+\frac{2(r-1)\cdot Y^{(b)}}{D+r-n-1}\cdot I^{(r-2)}_{n,\widehat{b}}),
			\label{eq:relation 01}
		\end{aligned}
	\end{equation}
	where the coefficients are
	
	\begin{equation}
		\begin{aligned}
			X^{(b)}=&\left((\overline{H_bL})(\overline{VL})_b-(\overline{H_bV})(\overline{LL})_b\right)/(\overline{LL}),\\
			Y^{(b)}=&\left((\overline{H_bL})R^2+(\overline{H_bV})(\overline{VL})_b-(\overline{H_bL})(\overline{VV})_b\right)/(\overline{LL}).
		\end{aligned}
  \label{eq:define XY}
	\end{equation}
	
	We focus on the summation of the tensor integrals with form
	
	\begin{equation}
		\begin{aligned}
			\phi_n(\mathcal{t})=&\sum_{r=0}^{\infty}\mathcal{t}^r\cdot I^{(r)}_n=\int\frac{d^Dl}{i\pi^{D/2}}\frac{1}{1-\mathcal{t}(2R\cdot l)}\frac{1}{\prod^n_{j=1}(l-q_j)^2-M_j^2},\\
			\phi_{n,\widehat{\textbf{b}}}(\mathcal{t})=&\sum_{r=0}^{\infty}\mathcal{t}^r\cdot I^{(r)}_{n,\widehat{\textbf{b}}}=\int\frac{d^Dl}{i\pi^{D/2}}\frac{1}{1-\mathcal{t}(2R\cdot l)}\frac{1}{\prod^n_{j=1,j\notin \textbf{b}}(l-q_j)^2-M_j^2}.
		\end{aligned}
	\end{equation}
	Then, we multiply both sides of equation \eqref{eq:relation 01} by $\mathcal{t}^r$ and sum over $r$ from $1$ to $\infty$. The following relations can be used:

	\begin{equation}
		\begin{aligned}
			&\sum_{r=1}^{\infty}\mathcal{t}^r\cdot I^{(r)}_n=\phi_n(\mathcal{t})-I^{(0)}_n,\\
			&\sum_{r=1}^{\infty}r\cdot \mathcal{t}^r\cdot I^{(r)}_n=\mathcal{t}\cdot \sum_{r=0}^{\infty}r\cdot \mathcal{t}^{(r-1)}\cdot I^{(r)}_n=\mathcal{t}\cdot \phi'_n(\mathcal{t}),\\
			&\sum_{r=1}^{\infty}\mathcal{t}^r\cdot I^{(r-1)}_n=\sum_{r=0}^{\infty}\mathcal{t}\cdot \mathcal{t}^r\cdot I^{(r)}_n=\mathcal{t}\cdot\phi_n(\mathcal{t}),\\
			&\sum_{r=1}^{\infty}r\cdot \mathcal{t}^r\cdot I^{(r-1)}_n=\sum_{r=0}^{\infty}(r+1)\cdot \mathcal{t}\cdot \mathcal{t}^r\cdot I^{(r)}_n=\mathcal{t}^2\cdot \phi'_n(\mathcal{t})+\mathcal{t}\cdot \phi_n (\mathcal{t}),\\
			&\sum_{r=2}^{\infty}\mathcal{t}^r\cdot I^{(r-2)}_n=\sum_{r=0}^{\infty}\mathcal{t}^{(r+2)}\cdot I^{(r)}_n=\mathcal{t}^2\cdot \phi_n(\mathcal{t}),\\
			&\sum_{r=2}^{\infty}r\cdot \mathcal{t}^r\cdot I^{(r-2)}_n=\sum_{r=0}^{\infty}(r+2)\cdot \mathcal{t}^{(r+2)}\cdot I^{(r)}_n=\mathcal{t}^3\phi'_n(\mathcal{t})+2\mathcal{t}^2\phi_n(\mathcal{t}).    
		\end{aligned}
	\end{equation}
	We obtain a differential function  for the generating function $\phi_n(\mathcal{t})$ :
	\begin{equation}
		\begin{aligned}
                &\bigg((D-n-1)-2(D-n)\cdot\frac{(\overline{VL})}{(\overline{LL})}\mathcal{t}-4\cdot\frac{R^2-(\overline{VV})}{(\overline{LL})}\mathcal{t}^2\bigg)\phi _n(\mathcal{t})\\
			+&\bigg(\mathcal{t}-4\cdot\frac{(\overline{VL})}{(\overline{LL})}\mathcal{t}^2-4\cdot\frac{R^2-(\overline{VV})}{(\overline{LL})}\mathcal{t}^3\bigg)\phi'_n(\mathcal{t})-(D-n-1)I^{(0)}_n\\
			=&\sum_{b=1}^n \Bigg\{X^{(b)}\bigg((D-n)\mathcal{t}\cdot\phi_{n;\widehat{b}}(\mathcal{t})+\mathcal{t}^2\cdot\phi'_{n;\widehat{b}}(\mathcal{t})\bigg)+2Y^{(b)}\bigg(\mathcal{t}^3\cdot\phi'_{n;\widehat{b}}(\mathcal{t})+\mathcal{t}^2\cdot\phi_{n;\widehat{b}}(\mathcal{t})\bigg)\Bigg\}.
		\end{aligned}
		\label{eq:relation 001}
	\end{equation}

We know that any integral can be written as the linear combination of certain
irreducible scalar integrals (which are the master integrals in arbitrary spacetime dimension) with coefficients as the rational functions of external momenta,
masses and space-time dimension $D$. 

\begin{equation}
\begin{aligned}
&I_{n}^{(r)}=\sum_{\textbf{b}\subseteq \{1,2,...,n\}} C^{(r)}_{n\to n;\widehat{\textbf{b}}}\cdot    I_{n;\widehat{\textbf{b}}}^{(0)}\ ,\\ 
&\phi_n(\mathcal{t})=\sum_{\textbf{b}\subseteq \{1,2,...,n\}} \Big\{\sum_{r=0}^\infty \mathcal{t}^r\cdot C^{(r)}_{n\to n;\widehat{\textbf{b}}}\Big\}\cdot I_{n;\widehat{\textbf{b}}}^{(0)}=\sum_{\textbf{b}\subseteq\{1,2,...,n\}}\textbf{GF}_{n\to n;\widehat{\textbf{b}}}(\mathcal{t})\cdot I_{n;\widehat{\textbf{b}}}^{(0)}\ .
\end{aligned}
\end{equation}
Thus, equation  \eqref{eq:relation 001} is transformed into a recursive formula for the generating functions of the reduction coefficients.

\begin{equation}
    \begin{aligned}
        &\bigg((D-n-1)-2(D-n)\cdot\frac{(\overline{VL})}{(\overline{LL})}\mathcal{t}-4\cdot\frac{R^2-(\overline{VV})}{(\overline{LL})}\mathcal{t}^2\bigg)\textbf{GF}_{n\to n;\widehat{\textbf{b}}}(\mathcal{t})\\
			&+\bigg(\mathcal{t}-4\cdot\frac{(\overline{VL})}{(\overline{LL})}\mathcal{t}^2-4\cdot\frac{R^2-(\overline{VV})}{(\overline{LL})}\mathcal{t}^3\bigg)\textbf{GF}'_{n\to n;\widehat{\textbf{b}}}(\mathcal{t})\\
			=&\sum_{b_i\in \textbf{b}} \Bigg\{X^{(b_i)}\bigg((D-n)\mathcal{t}\cdot\textbf{GF}_{n;\widehat{b_i}\to n;\widehat{\textbf{b}}}(\mathcal{t})+\mathcal{t}^2\cdot \textbf{GF}'_{n;\widehat{b_i}\to n;\widehat{\textbf{b}}}(\mathcal{t})\bigg)\\
   &+2Y^{(b_i)}\bigg(\mathcal{t}^3\cdot \textbf{GF}'_{n;\widehat{b_i}\to n;\widehat{\textbf{b}}}(\mathcal{t})+\mathcal{t}^2\cdot\textbf{GF}_{n;\widehat{b_i}\to n;\widehat{\textbf{b}}}(\mathcal{t})\bigg)\Bigg\}+(D-n-1)C^{(0)}_{n\to n;\widehat{\textbf{b}}}\ ,
    \end{aligned}
    \label{eq:relation 02}
\end{equation}
where $C^{(0)}_{n\to n;\widehat{\textbf{b}}}$ is the reduction coefficient of scalar integral $I^{(0)}_{n}$ to Master Integral $I^{(0)}_{n;\widehat{\textbf{b}}}$. Because $I^{(0)}_{n}$ is an irreducible scalar integral, the reduction coefficient $ C^{(0)}_{n\to n;\widehat{\textbf{b}}} $ is 1 if $ \textbf{b}$ is the null set, i.e, $\textbf{b}=\emptyset$ and  $ C^{(0)}_{n\to n;\widehat{\textbf{b}}} =0$ otherwise. $\textbf{GF}_{n;\widehat{b_i}\to n;\widehat{\textbf{b}}}(\mathcal{t})$ in the right hand side of \eqref{eq:relation 02} is the generating function for the reduction coefficient of $\phi_{n;\widehat{b_i}}(\mathcal{t})$ to Master Integral $I^{(0)}_{n;\widehat{\textbf{b}}}$.	The equation \eqref{eq:relation 02} serves as the cornerstone of this paper, drawing a connection between the generating functions in the reduction of an $n$-gon to an $(n-k)$-gon and an $(n-1)$-gon to an $((n-1)-(k-1))$-gon. At first, if we set $\textbf{b}=\emptyset$, the terms inside the curly braces on the right-hand side of \eqref{eq:relation 02} vanish. The remaining part is a first-order non-homogeneous ordinary differential equation with respect to the generating function $\textbf{GF}_{n\to n}(\mathcal{\mathcal{t}})$. By solving this differential equation, we obtain analytic expression of the generating function for an $n$-gon to an $n$-gon. Once we have obtained the analytic expression for the generating function for the reduction of an $n$-gon to $(n-(k-1))$-gon, due to the arbitrariness of $n$, we can also know the analytic expression for an $(n-1)$-gon to $((n-1)-(k-1))$-gon (which is what appears on the right side of the equation). Then, the equation \eqref{eq:relation 02} becomes a first-order non-homogeneous differential equation for the generating function of an $n$-gon to $(n-k)$-gon. Continuously repeating this process, we can calculate all the generating functions.

	\section{$n$-gon to $n$-gon}
	\label{sec:n to n}
	For the case of reducing an $n$-gon to an $n$-gon, that is, $\textbf{b}=\emptyset$, we have, $C^{(0)}_{n\to n}=1$ and the terms inside the curly braces on the right-hand side of equation \eqref{eq:relation 02} vanish. Consequently, the differential equation for the generating function $\textbf{GF}_{n\to n}(\mathcal{t})$ simplifies to:

	\begin{equation}
		\begin{aligned}
			&\left((D-n-1)-2(D-n)\cdot\frac{(\overline{VL})}{(\overline{LL})}\mathcal{t}-4\cdot\frac{R^2-(\overline{VV})}{(\overline{LL})}\mathcal{t}^2\right)\textbf{GF}_{n\to n}(\mathcal{t})\\
			+&\left(\mathcal{t}-4\cdot\frac{(\overline{VL})}{(\overline{LL})}\mathcal{t}^2-4\cdot\frac{R^2-(\overline{VV})}{(\overline{LL})}\mathcal{t}^3\right)\textbf{GF}'_{n\to n}(\mathcal{t})=(D-n-1).
		\end{aligned}
	\end{equation}
	
	The general solution to this differential equation is\footnote{See Appendix \ref{sec:int}.}:
 
	\begin{equation}
		\begin{aligned}
			&\mathcal{t}^{1-D+n}\left(1-x_{+}\cdot \mathcal{t}\right)^{\frac{D-n-2}{2}}\left(1-x_{-}\cdot \mathcal{t}\right)^{\frac{D-n-2}{2}}\cdot C_1+\frac{1}{1-x_{+}\cdot \mathcal{t}}\cdot \ _2F_1\left(\begin{array}{l}
1 , \frac{D-n}{2}  \\
 D-n 
\end{array} \bigg| \frac{(x_{-} - x_{+})\cdot \mathcal{t}}{1 - x_+\cdot \mathcal{t}}\right),
		\end{aligned}
	\end{equation}
 where 
\begin{equation}
\begin{aligned}
x_{\pm}&=\frac{2\bigg((\overline{VL})\pm\sqrt{(\overline{LL})R^2+(\overline{VL})^2-(\overline{LL})(\overline{VV})}\bigg)}{(\overline{LL})}.
    \end{aligned}
    \label{eq:xpm n to n}
\end{equation}
The Generalized Hypergeometric Function is defined as:
	\begin{equation}
_pF_q\left(\begin{array}{l}
a_1 , a_2 , \cdots , a_p \\
b_1 , b_2 , \cdots , b_q
\end{array} \bigg| z\right)=\sum_{n=0}^\infty\frac{(a_1)_n\cdots(a_p)_n}{(b_1)_n\cdots (b_q)_n}\cdot \frac{z^n}{n!},
\end{equation}
where we have used the \textbf{Pochhammer symbol}

\begin{equation}
(x)_n\equiv \frac{\Gamma(x+n)}{\Gamma(x)}=\prod_{i=1}^n(x+(i-1)).
	\end{equation}

At first glance, one might expect the undetermined constant to be resolved by the initial condition $\textbf{GF}_{n\to n}(0)=C^{(0)}_{n\to n}=1$. However, this doesn't work in our case. So, how should we determine $C_1$? The definition of $\textbf{GF}_{n\to n}(\mathcal{t})$ specifies that the generating functions should be represented as a Taylor series of $\mathcal{t}$. Nevertheless, the term 
\begin{equation}
\mathcal{t}^{1-D+n}\left(1-x_{+}\cdot \mathcal{t}\right)^{\frac{D-n-2}{2}}\left(1-x_{-}\cdot \mathcal{t}\right)^{\frac{D-n-2}{2}},
\end{equation}
cannot be expanded into a Taylor series\footnote{Typically, $D$ is not selected as an integer due to dimensional regularization} of $\mathcal{t}$. Thus, the constant $C_1=0$. Ultimately, we obtain the generating function of the reduction from an $n$-gon to an $n$-gon as follows:
 
	\begin{equation}
		\textbf{GF}_{n\to n}(\mathcal{t})=\frac{1}{1-x_{+}\cdot \mathcal{t}}\cdot \ _2F_1\left(\begin{array}{l}
1 , \frac{D-n}{2}  \\
 D-n 
\end{array} \bigg| \frac{(x_{-} - x_{+})\mathcal{t}}{1 - x_+\cdot \mathcal{t}}\right).
		\label{eq:n to n}
	\end{equation}

	\subsection{Analytical Results}
Upon deriving the formula for the generating function pertaining to the reduction from an $n$-gon to an $n$-gon, as delineated in equation \eqref{eq:n to n}, we subsequently elucidate the corresponding analytical formalism.

Substituting \eqref{eq:xpm n to n} into \eqref{eq:n to n}, we can obtain:
\begin{equation}
		\textbf{GF}_{n\to n}(\mathcal{t})=\frac{1}{1-\frac{2[(\overline{VL})+\sqrt{(\overline{LL})R^2+(\overline{VL})^2-(\overline{LL})(\overline{VV})}]}{(\overline{LL})}\cdot \mathcal{t}}\cdot \ _2F_1\left(\begin{array}{l}
1 , \frac{D-n}{2}  \\
 D-n 
\end{array} \bigg| \mathcal{x}(\mathcal{t)}\right),
	\end{equation}
 where
 \begin{equation}
		\begin{aligned} \mathcal{x}(\mathcal{t})=-\frac{4\sqrt{(\overline{LL})R^2+(\overline{VL})^2-(\overline{LL})(\overline{VV})} \cdot\mathcal{t}}{(\overline{LL})-2[(\overline{VL})+\sqrt{(\overline{LL})R^2+(\overline{VL})^2-(\overline{LL})(\overline{VV})}]\cdot\mathcal{t}},
		\end{aligned}
	\end{equation}
and
\begin{equation}
\begin{aligned}
    &(\overline{LL})=L\cdot Q^{-1}\cdot L\\
    &=\begin{pmatrix}1&1&\cdots&1 \end{pmatrix}\cdot \begin{pmatrix}M_1^2&\frac{M^2_2+M^2_1-(q_2-q_1)^2}{2}&\cdots&\frac{M^2_1+M^2_n-(q_1-q_n)^2}{2}\\\frac{M^2_2+M^2_1-(q_2-q_1)^2}{2}&M_2^2&\cdots &\frac{M^2_2+M^2_n-(q_2-q_n)^2}{2}\\ \vdots&\vdots&\ddots&\vdots\\
    \frac{M^2_n+M^2_1-(q_n-q_1)^2}{2}&\frac{M^2_n+M^2_2-(q_n-q_2)^2}{2} &\cdots&M^2_n
    \end{pmatrix}^{-1}\cdot \begin{pmatrix}1\\1\\\vdots\\1 \end{pmatrix},\\
 &(\overline{VL})=V\cdot Q^{-1}\cdot L\\
 &=\begin{pmatrix}R\cdot q_1&R\cdot q_2&\cdots&R\cdot q_n \end{pmatrix}\cdot \begin{pmatrix}M_1^2&\frac{M^2_2+M^2_1-(q_2-q_1)^2}{2}&\cdots&\frac{M^2_1+M^2_n-(q_1-q_n)^2}{2}\\\frac{M^2_2+M^2_1-(q_2-q_1)^2}{2}&M_2^2&\cdots &\frac{M^2_2+M^2_n-(q_2-q_n)^2}{2}\\ \vdots&\vdots&\ddots&\vdots\\
    \frac{M^2_n+M^2_1-(q_n-q_1)^2}{2}&\frac{M^2_n+M^2_2-(q_n-q_2)^2}{2} &\cdots&M^2_n
    \end{pmatrix}^{-1}\cdot \begin{pmatrix}1\\1\\\vdots\\1 \end{pmatrix},\\
 &(\overline{VV})=V\cdot Q^{-1}\cdot V\\
 &=\begin{pmatrix}R\cdot q_1&R\cdot q_2&\cdots&R\cdot q_n \end{pmatrix}\cdot \begin{pmatrix}M_1^2&\frac{M^2_2+M^2_1-(q_2-q_1)^2}{2}&\cdots&\frac{M^2_1+M^2_n-(q_1-q_n)^2}{2}\\\frac{M^2_2+M^2_1-(q_2-q_1)^2}{2}&M_2^2&\cdots &\frac{M^2_2+M^2_n-(q_2-q_n)^2}{2}\\ \vdots&\vdots&\ddots&\vdots\\
    \frac{M^2_n+M^2_1-(q_n-q_1)^2}{2}&\frac{M^2_n+M^2_2-(q_n-q_2)^2}{2} &\cdots&M^2_n
    \end{pmatrix}^{-1} \cdot \begin{pmatrix}R\cdot q_1\\R\cdot q_2\\\vdots\\R\cdot q_n .\end{pmatrix}.
    \end{aligned}
\end{equation}
By expanding the generating function as a series in terms of $\mathcal{t}$, we obtain:

\begin{equation}
\textbf{GF}_{n\to n}(\mathcal{t})=\sum_{r=0}^\infty \mathcal{t}^r\cdot C^{(r)}_{n\to n},
\end{equation}
where $ C^{(r)}_{n\to n}$ are the reduction coefficients of tensor integral
\begin{equation}
		I^{(r)}_n=\int\frac{d^Dl}{i\pi^{D/2}}\frac{(2R\cdot l)^r}{\prod^n_{j=1}(l-q_j)^2-M_j^2},
	\end{equation}
to the irreducible scalar integral (Master Integral)\  $I_{n}^{(0)}$. Furthermore, using the Taylor series expansion,
\begin{equation}
     _pF_q\left(\begin{array}{l}
a_1 , a_2 , \cdots , a_p \\
b_1 , b_2 , \cdots , b_q
\end{array} \bigg| \frac{m\cdot z}{1-n\cdot z}\right)=1+\sum_{k=1}^{\infty}\Bigg(\sum_{i=1}^k\frac{\textbf{C}^{i-1}_{k-1}(a_1)_i(a_2)_i\cdots (a_p)_im^in^{k-i}}{i!(b_1)_i(b_2)_i\cdots (b_q)_i}\Bigg)z^k,
\label{eq:Taylor 01}
\end{equation}
where $\textbf{C}^{i-1}_{k-1}$ are binomial coefficients defined as $\textbf{C}^{i-1}_{k-1}=\frac{(k-1)!}{(i-1)!(k-1-(i-1))!}$, we can obtain the coefficients for each order in the series expansion:
  \begin{equation}
C_{n \to n}^{(r)}=x_{+}^r+\sum_{j=1}^{r}\sum_{i=1}^{j} \bigg( \frac{(\frac{D-n}{2})_{i} }{(D-n)_{i}}\big(x_{-}-x_{+}\big)^{i}x_{+}^{r-i}\textbf{C}^{i-1}_{j-1}\bigg).
\end{equation}

	\section{$n$-gon to $(n-1)$-gon}
    \label{sec:n to n-1}
	
	Next, let us consider the case of the $(n-1)$-gon. Without loss of generality, we select the Master Integral as follows:

\begin{equation}
I^{(0)}_{n,\widehat{a_1}}=\int\frac{d^Dl}{i\pi^{D/2}}\frac{1}{\prod^n_{j=1,j\neq a_1}(l-q_j)^2-M_j^2}.
\end{equation}
In this case, $C^{(0)}_{n\to n;\widehat{a_1}}=0$. Then equation \eqref{eq:relation 02} transforms into:

	\begin{equation}
		\begin{aligned}
			&\left((D-n-1)-2(D-n)\cdot\frac{(\overline{VL})}{(\overline{LL})}\mathcal{t}-4\cdot\frac{R^2-(\overline{VV})}{(\overline{LL})}\mathcal{t}^2\right)\textbf{GF}_{n\to n;\widehat{a_1}}(\mathcal{t})\\
			+&\left(\mathcal{t}-4\cdot\frac{(\overline{VL})}{(\overline{LL})}\mathcal{t}^2-4\cdot\frac{R^2-(\overline{VV})}{(\overline{LL})}\mathcal{t}^3\right)\textbf{GF}'_{n\to n;\widehat{a_1}}(\mathcal{t})\\
			=&X^{(a_1)}\bigg((D-n)\mathcal{t}\cdot\textbf{GF}_{n;\widehat{a_1}\to n;\widehat{a_1}}(\mathcal{t})+\mathcal{t}^2\cdot\textbf{GF}'_{n;\widehat{a_1}\to n;\widehat{a_1}}(\mathcal{t})\bigg)\\
   +&2Y^{(a_1)}\bigg(\mathcal{t}^3\cdot\textbf{GF}'_{n;\widehat{a_1}\to n;\widehat{a_1}}(\mathcal{t})+\mathcal{t}^2\cdot\textbf{GF}_{n;\widehat{a_1}\to n;\widehat{a_1}}(\mathcal{t})\bigg),
		\end{aligned}
		\label{eq:dif eq n to n-1}
	\end{equation}
where $\textbf{GF}_{n;\widehat{a_1}\to n;\widehat{a_1}}(\mathcal{t})$ can be obtained from \eqref{eq:n to n}. All we need to do is replace all instances of $n$ with $(n-1)$ in expression \eqref{eq:n to n}, and replace all instances of $(\overline{LL}),(\overline{VL}),(\overline{VV})$ with $(\overline{LL})_{a_1},(\overline{VL})_{a_1},(\overline{VV})_{a_1}$ respectively. That is,   $\textbf{GF}_{n;\widehat{a_1}\to n;\widehat{a_1}}(\mathcal{t})=[\textbf{GF}_{n-1\to n-1}(\mathcal{t})]_{a_1}$,
	
	\begin{equation}
		\begin{aligned}
			\textbf{GF}_{n,\widehat{a_1}\to n,\widehat{a_1}}(\mathcal{t})=&\frac{1}{1-[x_{+}]_{a_1}\cdot \mathcal{t}}\cdot \ _2F_1\left(\begin{array}{l}
1 , \frac{D-n+1}{2}  \\
 D-n+1 
\end{array} \bigg| \frac{[x_{-} - x_{+}]_{a_1}\cdot \mathcal{t}}{1 - [x_+]_{a_1}\cdot \mathcal{t}}\right),
\label{eq:n-1 to n-1}
   \end{aligned}
  \end{equation}
where the notation
   \begin{equation}
\begin{aligned}
[x_{\pm}]_{a_1}&=\frac{2\bigg((\overline{VL})_{a_1}\pm\sqrt{(\overline{LL})_{a_1}R^2+(\overline{VL})_{a_1}^2-(\overline{LL})_{a_1}(\overline{VV})_{a_1}}\bigg)}{(\overline{LL})_{a_1}},
    \end{aligned}
\end{equation}
as defined in  \Autoref{sec:prep}. 
	The general solution to equation \eqref{eq:dif eq n to n-1} is\footnote{See Appendix \ref{sec:int}.}

	\begin{equation}
		\begin{aligned}
			&\mathcal{t}^{1-D+n}(1-x_{+}\cdot\mathcal{t})^{\frac{-2+D-n}{2}}(1-x_{-}\cdot\mathcal{t})^{\frac{-2+D-n}{2}}\\
			\times & \Bigg\{C_1+\int_0^\mathcal{t}\mathcal{u}^{-1+D-n}(1-x_{+}\cdot \mathcal{u})^{\frac{n-D}{2}}(1-x_{-}\cdot\mathcal{u})^{\frac{n-D}{2}}\\
&\times\Bigg(2Y^{(a_1)}\big(\mathcal{u}^2\cdot\textbf{GF}'_{n;\widehat{a_1}\to n;\widehat{a_1}}(\mathcal{u})+\mathcal{u}\cdot\textbf{GF}_{n;\widehat{a_1}\to n;\widehat{a_1}}(\mathcal{u})\big)\\
			&+X^{(a_1)}\big((D-n)\textbf{GF}_{n;\widehat{a_1}\to n;\widehat{a_1}}(\mathcal{u})+\mathcal{u}\cdot\textbf{GF}'_{n;\widehat{a_1}\to n;\widehat{a_1}}(\mathcal{u})\big)\Bigg)d\mathcal{u} \Bigg\}.
		\end{aligned}
  \label{eq:GS n to n-1 01}
	\end{equation}
	Clearly, the initial condition $\textbf{GF}_{n\to n,\widehat{a_1}}(0)=0$ is insufficient to determine the undetermined coefficient $C_1$. However, since we require the generating function to be a Taylor series of $\mathcal{t}$, we must have $C_1=0$.\footnote{$(1-x_{+}\cdot \mathcal{u})^{\frac{n-D}{2}}(1-x_{-}\cdot\mathcal{u})^{\frac{n-D}{2}}$ and the terms in the large square brackets can be expanded as a Taylor series of $\mathcal{u}$. The factor $\mathcal{u}^{-1+D-n}$ in the integrand will cancel the factor $\mathcal{t}^{1-D+n}$ outside the integrand. This makes the result a Taylor series of $\mathcal{t}$ after integrating.}

The integral in \eqref{eq:GS n to n-1 01} cannot be evaluated directly. An intuitive idea would be to first expand the integrand into a series in terms of $\mathcal{u}$, and then integrate. However, fully expanding a hypergeometric function into a series would render the expression overly complex. Therefore, we decide to select a suitable integral formula and perform a series expansion only on part of the integrand. The integral formula we select is

\begin{equation}
		\begin{aligned}
			&\int dz\ _pF_q\left(\begin{array}{l}
a_1,...,a_p  \\
 b_1,...,b_q 
\end{array} \bigg| k\cdot z\right)\cdot z^m=\frac{z^{m+1}}{m+1}\cdot\  _{(p+1)}F_{(q+1)}\left(\begin{array}{l}
a_1, ..., a_p, 1+m  \\
b_1, ..., b_q, 2+m
\end{array} \bigg| k\cdot z\right)+Const.
		\end{aligned}
		\label{eq:formula 02}
	\end{equation}
 Then in order to evaluate the integral explicitly, we need to change the integration variable 
	
	\begin{equation}
		\mathcal{u}\to \mathcal{x}_{a_1}(\mathcal{u})=\frac{\mathcal{u}}{1 - [x_+]_{a_1}\cdot \mathcal{u}}.
	\end{equation}
	Note that in \eqref{eq:GS n to n-1 01} there are terms involving the derivative of the generating function $\textbf{GF}_{n;\widehat{a_1}\to n;\widehat{a_1}}$, which requires the use of the following derivative formula for the generalized hypergeometric function

	\begin{equation}
		\frac{d}{dz}\ _pF_q\left(\begin{array}{l}
a_1,...,a_p  \\
 b_1,...,b_q 
\end{array} \bigg|z\right)=\frac{\prod_{i=1}^p a_i}{\prod_{j=1}^q b_j}\cdot\ _pF_q\left(\begin{array}{l}
1+a_1,...,1+a_p  \\
 1+b_1,...,1+b_q 
\end{array} \bigg|z\right).
	\end{equation}
	Substituting \eqref{eq:n-1 to n-1} in \eqref{eq:GS n to n-1 01}, after altering the integration variable, the integral part becomes
	
	\begin{equation}
		\begin{aligned}
			&\int_0^{\mathcal{x}_{a_1}(\mathcal{t})}x^{D-n-1}\Bigg\{(1+\textbf{A}_{a_1}x+\textbf{B}_{a_1}x^2)^{\frac{n-D}{2}}(\textbf{P}^{(0)}_{a_1}+\textbf{P}^{(1)}_{a_1}x)\cdot \ _2F_1\left(\begin{array}{l}
1 , \frac{D-n+1}{2}  \\
 D-n+1 
\end{array} \bigg|[x_{-}-x_{+}]_{a_1}\cdot x\right)\\
			&+(1+\textbf{A}_{a_1}x+\textbf{B}_{a_1}x^2)^{\frac{n-D}{2}}(\textbf{Q}^{(1)}_{a_1}x+\textbf{Q}^{(2)}_{a_1}x^2)\cdot \ _2F_1\left(\begin{array}{l}
2 , \frac{D-n+3}{2}  \\
 D-n+2 
\end{array} \bigg|[x_{-}-x_{+}]_{a_1}\cdot x\right)\Bigg\}dx,
		\end{aligned}
	\end{equation}
	where 
	
	\begin{equation}
		\begin{aligned}
			&\textbf{A}_{a_1}=-(x_{+}+x_{-})+2[x_{+}]_{a_1},\\
			&\textbf{B}_{a_1}=x_{+}\cdot x_{-}-(x_{+}+x_{-})\cdot [x_{+}]_{a_1}+[x_{+}^2]_{a_1},\\
   &\textbf{P}^{(0)}_{a_1}(n)=(D-n)\cdot X^{(a_1)},\\&\textbf{P}^{(1)}_{a_1}=2Y^{(a_1)}+[x_{+}]_{a_1}\cdot X^{(a_1)},\\
			&\textbf{Q}^{(1)}_{a_1}= -[x_{+}-x_{-}]_{a_1}\cdot X^{(a_1)}/2,\\
            &\textbf{Q}^{(2)}_{a_1}=-\frac{1}{2}\cdot (2Y^{(a_1)}+[x_{+}]_{a_1}\cdot X^{(a_1)})\cdot [x_{+}-x_{-}]_{a_1}.
		\end{aligned}
  \label{eq: define abpq}
	\end{equation}
	We expand 
	
	\begin{equation}
		\begin{aligned}
			&(1+\textbf{A}_{a_1}x+\textbf{B}_{a_1}x^2)^{\frac{n-D}{2}}(\textbf{P}^{(0)}_{a_1}+\textbf{P}^{(1)}_{a_1}x)=\sum_{m=0}^\infty \textbf{C}^{(1)}_{a_1}(n;m)\cdot  x^m,\\
			&(1+\textbf{A}_{a_1}x+\textbf{B}_{a_1}x^2)^{\frac{n-D}{2}}(\textbf{Q}^{(1)}_{a_1}x+\textbf{Q}^{(2)}_{a_1}x^2)=\sum_{m=0}^\infty \textbf{C}^{(2)}_{a_1}(n;m)\cdot  x^m,
		\end{aligned}
	\end{equation}
 as a Taylor series of $x$. By formula 
	
	\begin{equation}
		(1+\textbf{A} x+\textbf{B} x^2)^c=\sum_{m=0}^\infty \Big\{\sum_{i=0}^{\lfloor\frac{m}{2}\rfloor}\frac{(c-m+i+1)_{(m-i)}}{(m-2i)!\cdot i!}\textbf{A}^{m-2i}\textbf{B}^i\Big\} x^m,
		\label{eq:formula 01}
	\end{equation}
we have the expansion coefficients as follows:

	\begin{equation}
		\begin{aligned}
			\textbf{C}_{a_1}^{(1)}(n;m)=&\textbf{P}_{a_1}^{(0)}(n)\cdot \sum_{i=0}^{\lfloor\frac{m}{2}\rfloor}\frac{(\frac{n-D}{2}-m+i+1)_{(m-i)}}{(m-2i)!\cdot i!}\textbf{A}_{a_1}^{m-2i}\textbf{B}_{a_1}^i\\
			+&\textbf{P}_{a_1}^{(1)}\cdot \sum_{i=0}^{\lfloor\frac{m-1}{2}\rfloor}\frac{(\frac{n-D}{2}-m+i+2)_{(m-1-i)}}{(m-1-2i)!\cdot i!}\textbf{A}_{a_1}^{m-1-2i}\textbf{B}_{a_1}^i,\\
			\textbf{C}_{a_1}^{(2)}(n;m)=
			&\textbf{Q}_{a_1}^{(1)}\cdot \sum_{i=0}^{\lfloor\frac{m-1}{2}\rfloor}\frac{(\frac{n-D}{2}-m+i+2)_{(m-1-i)}}{(m-1-2i)!\cdot i!}\textbf{A}_{a_1}^{m-1-2i}\textbf{B}_{a_1}^i\\
			+&\textbf{Q}_{a_1}^{(2)}\cdot \sum_{i=0}^{\lfloor\frac{m-2}{2}\rfloor}\frac{(\frac{n-D}{2}-m+i+3)_{(m-2-i)}}{(m-2-2i)!\cdot i!}\textbf{A}_{a_1}^{m-2-2i}\textbf{B}_{a_1}^i.
		\end{aligned}
  \label{eq:c1 and c2}
	\end{equation}
	Then the integral part (excluding the constant) becomes:

	\begin{equation}
		\begin{aligned}
			&\sum_{m=0}^\infty \textbf{C}^{(1)}_{a_1}(n;m)\int_0^{\mathcal{x}_{a_1}(\mathcal{t})}dx\  x^{D-n-1+m}\cdot \ _2F_1\left(\begin{array}{l}
1 , \frac{D-n+1}{2}  \\
 D-n+1 
\end{array} \bigg|[x_{-}-x_{+}]_{a_1}\cdot x\right)\\
			+&\sum_{m=0}^\infty \textbf{C}^{(2)}_{a_1}(n;m)\int_0^{\mathcal{x}_{a_1}(\mathcal{t})}dx\  x^{D-n-1+m}\cdot \ _2F_1\left(\begin{array}{l}
2 , \frac{D-n+3}{2}  \\
 D-n+2 
\end{array} \bigg|[x_{-}-x_{+}]_{a_1}\cdot x\right).
		\end{aligned}
	\end{equation}
	By using integral formula of generalized hypergeometric function\eqref{eq:formula 02}, we can explicitly solve for the generating function of the reduction from an $n$-gon to an $(n-1)$-gon,
\begin{equation}
		\begin{aligned}
			&\textbf{GF}_{n\to n;\widehat{a_1}}(\mathcal{t})\\=&\left((1-x_{+}\cdot \mathcal{t})(1-x_{-}\cdot \mathcal{t})\right)^{\frac{-2+D-n}{2}}\times \sum_{m=0}^{\infty}\Bigg\{\frac{\mathcal{t}^{m+1}}{(m+D-n)(1-[x_{+}]_{a_1}\cdot \mathcal{t})^{m+D-n}}\\
			&\times \bigg\{\textbf{C}_{a_1}^{(1)}(n;m)\cdot\  _3F_2\left(\begin{array}{l}
1,\frac{D-n+1}{2},m+D-n  \\
 D-n+1,m+D-n+1
\end{array} \bigg|\frac{[x_{-}-x_{+}]_{a_1}\cdot\mathcal{t}}{1 - [x_+]_{a_1}\cdot \mathcal{t}}\right)\\
			&\ \ +\textbf{C}_{a_1}^{(2)}(n;m)\cdot\  _3F_2\left(\begin{array}{l}
2,\frac{D-n+3}{2},m+D-n  \\
 D-n+2,m+D-n+1
\end{array} \bigg|\frac{[x_{-}-x_{+}]_{a_1}\cdot\mathcal{t}}{1 - [x_+]_{a_1}\cdot \mathcal{t}}\right)\bigg\}\Bigg\}.
		\end{aligned}
		\label{eq:GF n to n-1}
	\end{equation}

	\section{$n$-gon to $(n-2)$-gon}
 \label{sec:n to n-2}
	
	Now, let us discuss the reduction from an $n$-gon to an $(n-2)$-gon. Without loss of generality, we select the Master Integral:

	\begin{equation}
		I^{(0)}_{n,\widehat{a_1,a_2}}=\int \frac{d^Dl}{i\pi^\frac{D}{2}}\frac{1}{\prod_{j=1,j\neq a_1,a_2}^n (l-q_j)^2-M_j^2}.
	\end{equation}
	In this case, $C^{(0)}_{n\to n;\widehat{a_1,a_2}}=0$. Then recursive relation \eqref{eq:relation 02} transforms into 
	
		\begin{align}
			&\left((D-n-1)-2(D-n)\cdot\frac{(\overline{VL})}{(\overline{LL})}\mathcal{t}-4\cdot\frac{R^2-(\overline{VV})}{(\overline{LL})}\mathcal{t}^2\right)\textbf{GF}_{n\to n;\widehat{a_1,a_2}}(\mathcal{t})\notag\\
			+&\left(\mathcal{t}-4\cdot\frac{(\overline{VL})}{(\overline{LL})}\mathcal{t}^2-4\cdot\frac{R^2-(\overline{VV})}{(\overline{LL})}\mathcal{t}^3\right)\textbf{GF}'_{n\to n;\widehat{a_1,a_2}}(\mathcal{t})\notag\\
			=&\Bigg\{2Y^{(a_1)}\left(\mathcal{t}^3\cdot\textbf{GF}'_{n;\widehat{a_1}\to n;\widehat{a_1,a_2}}(\mathcal{t})+\mathcal{t}^2\cdot\textbf{GF}_{n;\widehat{a_1}\to n;\widehat{a_1,a_2}}(\mathcal{t})\right)\notag\\
			+&X^{(a_1)}\left((D-n)\mathcal{t}\cdot\textbf{GF}_{n;\widehat{a_1}\to n;\widehat{a_1,a_2}}(\mathcal{t})+\mathcal{t}^2\cdot\textbf{GF}'_{n;\widehat{a_1}\to n;\widehat{a_1,a_2}}(\mathcal{t})\right)\Bigg\}\notag\\
			+&\Bigg\{2Y^{(a_2)}\left(\mathcal{t}^3\cdot\textbf{GF}'_{n;\widehat{a_2}\to n;\widehat{a_1,a_2}}(\mathcal{t})+\mathcal{t}^2\cdot\textbf{GF}_{n;\widehat{a_2}\to n;\widehat{a_1,a_2}}(\mathcal{t})\right)\notag\\
			+&X^{(a_2)}\left((D-n)\mathcal{t}\cdot\textbf{GF}_{n;\widehat{a_2}\to n;\widehat{a_1,a_2}}(\mathcal{t})+\mathcal{t}^2\cdot\textbf{GF}'_{n;\widehat{a_2}\to n;\widehat{a_1,a_2}}(\mathcal{t})\right)\Bigg\}.\label{eq:dif eq n to n-2}
		\end{align}
	
	The general solution of \eqref{eq:dif eq n to n-2} is 
		\begin{align}
			&\mathcal{t}^{1-D+n}(1-x_{+}\cdot \mathcal{t})^{\frac{-2+D-n}{2}}(1-x_{-}\cdot \mathcal{t})^{\frac{-2+D-n}{2}}\notag\\
			\times & \Bigg\{C_1+\int_0^\mathcal{t}\mathcal{u}^{-1+D-n}(1-x_{+}\cdot\mathcal{u})^{\frac{n-D}{2}}(1-x_{-}\cdot \mathcal{u})^{\frac{n-D}{2}}\notag\\
			&\times\Big\{2Y^{(a_1)}\left(\mathcal{u}^2\cdot\textbf{GF}'_{n;\widehat{a_1}\to n;\widehat{a_1,a_2}}(\mathcal{u})+\mathcal{u}\cdot\textbf{GF}_{n;\widehat{a_1}\to n;\widehat{a_1,a_2}}(\mathcal{u})\right)\notag\\
			+&X^{(a_1)}\left((D-n)\cdot\textbf{GF}_{n;\widehat{a_1}\to n;\widehat{a_1,a_2}}(\mathcal{u})+\mathcal{u}\cdot\textbf{GF}'_{n;\widehat{a_1}\to n;\widehat{a_1,a_2}}(\mathcal{u})\right)\notag\\
			+&2Y^{(a_2)}\left(\mathcal{u}^2\cdot\textbf{GF}'_{n;\widehat{a_2}\to n;\widehat{a_1,a_2}}(\mathcal{u})+\mathcal{u}\cdot\textbf{GF}_{n;\widehat{a_2}\to n;\widehat{a_1,a_2}}(\mathcal{u})\right)\notag\\
			+&X^{(a_2)}\left((D-n)\cdot\textbf{GF}_{n;\widehat{a_2}\to n;\widehat{a_1,a_2}}(\mathcal{u})+\mathcal{u}\cdot\textbf{GF}'_{n;\widehat{a_2}\to n;\widehat{a_1,a_2}}(\mathcal{u})\right)\Big\}d\mathcal{u}\Bigg\},\label{eq:GS n to n-2}
		\end{align}
	where $\textbf{GF}_{n;\widehat{a_1}\to n;\widehat{a_1,a_2}}(\mathcal{t})$ and $\textbf{GF}_{n;\widehat{a_2}\to n;\widehat{a_1,a_2}}(\mathcal{t})$ can be obtained from \eqref{eq:GF n to n-1} by 
		\begin{align}
			\textbf{GF}_{n;\widehat{a_1}\to n;\widehat{a_1,a_2}}(\mathcal{t})=&[\textbf{GF}_{n-1\to n-1;\widehat{a_2}}(\mathcal{t})]_{a_1},\notag\\
			\textbf{GF}_{n;\widehat{a_2}\to n;\widehat{a_1,a_2}}(\mathcal{t})=&[\textbf{GF}_{n-1\to n-1;\widehat{a_1}}(\mathcal{t})]_{a_2}.
		\end{align}
 We select the undetermined constant $C_1=0$ since the generating function should be a Taylor series of $\mathcal{t}$. We can see that in the integrand of expression \eqref{eq:GS n to n-2}, $a_1$ and $a_2$ are completely symmetric. We can calculate only one half and obtain the other half by exploiting this symmetry. Moreover, since we are going to use integral formula \eqref{eq:formula 02} to compute the integral, we can divide $\textbf{GF}_{n;\widehat{a_1}\to n;\widehat{a_1,a_2}}(\mathcal{t})$ and $\textbf{GF}_{n;\widehat{a_2}\to n;\widehat{a_1,a_2}}(\mathcal{t})$ according to the generalized hypergeometric function as follows:
	\begin{equation}
		\begin{aligned}
			\textbf{GF}_{n;\widehat{a_1}\to n;\widehat{a_1,a_2}}(\mathcal{t})=\sum_{m_1}^\infty \sum_{l=1,2}[C^{(l)}_{a_2}(n-1;m_1)]_{a_1} \textbf{GF}^{(l)(m_1)}_{n;\widehat{a_1}\to n;\widehat{a_1,a_2}}(\mathcal{t}),\\
			\textbf{GF}_{n;\widehat{a_2}\to n;\widehat{a_1,a_2}}(\mathcal{t})=\sum_{m_1}^\infty \sum_{l=1,2}[C^{(l)}_{a_1}(n-1;m_1)]_{a_2} \textbf{GF}^{(l)(m_1)}_{n;\widehat{a_2}\to n;\widehat{a_1,a_2}}(\mathcal{t}),
		\end{aligned}
	\end{equation}
	where 
		\begin{align}
			&\textbf{GF}^{(1)(m_1)}_{n;\widehat{a_1}\to n;\widehat{a_1,a_2}}(\mathcal{t})
	=\frac{1}{m_1+D-n+1}(1-[x_{+}]_{a_1}\cdot\mathcal{t})^{\frac{-1+D-n}{2}}(1-[x_{-}]_{a_1}\cdot\mathcal{t})^{\frac{-1+D-n}{2}}\notag\\
			\times &\Bigg\{\frac{\mathcal{t}^{m_1+1}}{(1+[x_{+}]_{a_1,a_2}\cdot \mathcal{t})^{m_1+D-n+1}}\cdot\  _3F_2\left(\begin{array}{l}
1,\frac{D-n+2}{2},m_1+D-n+1  \notag\\
 D-n+2,m_1+D-n+2
\end{array} \bigg|\frac{[x_{-}-x_{+}]_{a_1,a_2}\cdot\mathcal{t}}{1 - [x_+]_{a_1,a_2}\cdot \mathcal{t}}\right)\Bigg\},\notag\\
&\ \notag\\
			&\textbf{GF}^{(2)(m_1)}_{n;\widehat{a_1}\to n;\widehat{a_1,a_2}}(\mathcal{t})
			=\frac{1}{m_1+D-n+1}(1-[x_{+}]_{a_1}\cdot\mathcal{t})^{\frac{-1+D-n}{2}}(1-[x_{-}]_{a_1}\cdot\mathcal{t})^{\frac{-1+D-n}{2}}\notag\\
			\times &\Bigg\{\frac{\mathcal{t}^{m_1+1}}{(1+[x_{+}]_{a_1,a_2}\cdot \mathcal{t})^{m_1+D-n+1}}\cdot\  _3F_2\left(\begin{array}{l}
2,\frac{D-n+4}{2},m_1+D-n+1  \notag\\
 D-n+3,m_1+D-n+2
\end{array} \bigg|\frac{[x_{-}-x_{+}]_{a_1,a_2}\cdot\mathcal{t}}{1 - [x_+]_{a_1,a_2}\cdot \mathcal{t}}\right)\Bigg\},\notag\\
&\ \notag\\
			&\textbf{GF}^{(1)(m_1)}_{n;\widehat{a_2}\to n;\widehat{a_1,a_2}}(\mathcal{t})=\textbf{GF}^{(1)(m_1)}_{n;\widehat{a_1}\to n;\widehat{a_1,a_2}}(\mathcal{t})\Big|_{a_1 \leftrightarrow a_2},\notag\\
   &\ \notag\\
			&\textbf{GF}^{(2)(m_1)}_{n;\widehat{a_2}\to n;\widehat{a_1,a_2}}(\mathcal{t})=\textbf{GF}^{(2)(m_1)}_{n;\widehat{a_1}\to n;\widehat{a_1,a_2}}(\mathcal{t})\Big|_{a_1 \leftrightarrow a_2}.
		\end{align}
Each part involves only one generalized hypergeometric function. The constants $[C^{(1)}_{a_2}(n-1;m_1)]_{a_1}$, $[C^{(1)}_{a_1}(n-1;m_1)]_{a_2}$, $[C^{(2)}_{a_2}(n-1;m_1)]_{a_1}$, and $[C^{(2)}_{a_1}(n-1;m_1)]_{a_2}$ can be factored out of the integral sign. By this separation, the integral in equation \eqref{eq:GS n to n-2} can be divided into several parts. We can calculate each part separately and add it up. Next we will use the following part as an example to illustrate the process:

	\begin{equation}
		\begin{aligned}
			&\int_0^\mathcal{t}\mathcal{u}^{-1+D-n}(1-x_{+}\cdot\mathcal{u})^{\frac{n-D}{2}}(1-x_{-}\cdot \mathcal{u})^{\frac{n-D}{2}}\\
			&\times\Big\{2Y^{(a_1)}\left(\mathcal{u}^2\cdot\textbf{GF}'^{(1)(m_1)}_{n;\widehat{a_1}\to n;\widehat{a_1,a_2}}(\mathcal{u})+\mathcal{u}\cdot\textbf{GF}^{(1)(m_1)}_{n;\widehat{a_1}\to n;\widehat{a_1,a_2}}(\mathcal{u})\right)\\
			&+X^{(a_1)}\left((D-n)\cdot\textbf{GF}^{(1)(m_1)}_{n;\widehat{a_1}\to n;\widehat{a_1,a_2}}(\mathcal{u})+\mathcal{u}\cdot\textbf{GF}'^{(1)(m_1)}_{n;\widehat{a_1}\to n;\widehat{a_1,a_2}}(\mathcal{u})\right)\Big\}d\mathcal{u}.
		\end{aligned}
		\label{eq:n to n-2 01}
	\end{equation}

\textbf{The main difference} from  what we did in Section \ref{sec:n to n-1} is that we use another derivative formula of the generalized hypergeometric function

\begin{equation}
\begin{aligned}
&\frac{d}{dz}\ _pF_q\left(\begin{array}{l}
a_1,a_2,\cdots,a_{p-1},\alpha \\
b_1,b_2,\cdots,b_{q-1},\alpha+1
\end{array} \bigg|z\right)\\
=&\frac{\alpha}{z}\cdot\left(_{(p-1)}F_{(q-1)}\left(\begin{array}{l}
a_1,a_2,\cdots,a_{p-1}\\
b_1,b_2,\cdots,b_{q-1}
\end{array} \bigg|z\right)-\  _pF_q\left(\begin{array}{l}
a_1,a_2,\cdots,a_{p-1},\alpha \\
b_1,b_2,\cdots,b_{q-1},\alpha+1
\end{array} \bigg|z\right)\right).
\end{aligned}
\label{eq:dev rule 01}
\end{equation}
After changing the integration variable from $\mathcal{u}$ to

	\begin{equation}
		\mathcal{u}\to \mathcal{x}_{a_1,a_2}(\mathcal{u})=\frac{\mathcal{u}}{1 - [x_+]_{a_1,a_2}\cdot \mathcal{u}},
	\end{equation}
	the integral \eqref{eq:n to n-2 01} becomes 
	
	\begin{equation}
		\begin{aligned}
			&\int_0^{\mathcal{x}_{a_1,a_2}(\mathcal{t})} dx\ x^{D-n+m_1}\\
       &\Bigg\{\Big\{
			(1+\textbf{A}_1(a_1;a_2)x+\textbf{A}_2(a_1;a_2)x^2)^{-\frac{D-n}{2}}          (1+\textbf{B}_1(a_1;a_2)x+\textbf{B}_2(a_1;a_2)x^2)^{\frac{D-n-1}{2}}\\
			\times &\left(\textbf{Q}_0(a_1;a_2)+\textbf{Q}_1 (a_1;a_2)x\right)\cdot\ _2F_1\left(\begin{array}{l}
1,\frac{D-n+2}{2} \\
D-n+2
\end{array} \bigg|[x_{-}-x_{+}]_{a_1,a_2}\cdot x\right)\Big\}\\
+&\Big\{(1+\textbf{A}_1(a_1;a_2)x+\textbf{A}_2(a_1;a_2)x^2)^{-\frac{D-n}{2}}(1+\textbf{B}_1(a_1;a_2)x+\textbf{B}_2(a_1;a_2)x^2)^{\frac{D-n-3}{2}}\\
			\times &\frac{(D-n-1)\Big(\textbf{P}_1(a_1;a_2)x+\textbf{P}_2 (a_1;a_2)x^2\Big)}{2(D-n+1+m_1)}\cdot\ _3F_2\left(\begin{array}{l}
1,\frac{D-n+2}{2},m_1+D-n+1 \\
D-n+2,m_1+D-n+2
\end{array} \bigg|[x_{-}-x_{+}]_{a_1,a_2}\cdot x\right)\Big\}\Bigg\},
		\end{aligned}
		\label{eq:n to n-2 02}
	\end{equation}
	where coefficients are
	
	\begin{equation}
		\begin{aligned}
			&\textbf{A}_1(a_1;a_2)=-(x_{+}+x_{-})+2[x_{+}]_{a_1,a_2},\\
			&\textbf{A}_2(a_1;a_2)=[x^2_{+}]_{a_1,a_2}+x_{+}\cdot x_{-}-(x_{+}+x_{-})[x_{+}]_{a_1,a_2},\\
			&\textbf{B}_1(a_1;a_2)=-[x_{+}+x_{-}]_{a_2}+2[x_{+}]_{a_1,a_2},\\
			&\textbf{B}_2(a_1;a_2)=[x^2_{+}]_{a_1,a_2}+[x_{+}\cdot x_{-}]_{a_2}-[x_{+}+x_{-}]_{a_2}\cdot[x_{+}]_{a_1,a_2},\\
			&\textbf{P}_1(a_1;a_2)=-\Big([x_{+}+x_{-}]_{a_2}X^{(a_2)}+4Y^{(a_2)}\Big),	\\	
			&\textbf{P}_2(a_1;a_2)=2Y^{(a_2)}([x_{+}+x_{-}]_{a_2}-2[x_{+}]_{a_1,a_2})+X^{(a_2)}(2[x_{+}\cdot x_{-}]_{a_2}-[x_{+}+x_{-}]_{a_2}[x_{+}]_{a_1,a_2}),\\
                &\textbf{Q}_0(a_1;a_2)=X^{(a_2)},\\
                &\textbf{Q}_1(a_1;a_2)=2Y^{(a_2)}+[x_{+}]_{a_1,a_2}\cdot X^{(a_2)}.\\
		\end{aligned}
		\label{eq:define ABPQ}
	\end{equation}
	\textbf{Note that labels $a_1,a_2$ are no longer symmetric in the above coefficients}. By equation \eqref{eq:formula 01}, we can expand the integral \eqref{eq:n to n-2 02} as follows:

	\begin{equation}
		\begin{aligned}
			&\sum_{m_2=0}^\infty\int_0^{\mathcal{x}_{a_1,a_2}(\mathcal{t})} x^{m_1+m_2+D-n}\Big\{(\textbf{0})_{m_2}(n;a_1;a_2;m_1)\cdot \ _2F_1\left(\begin{array}{l}
1,\frac{D-n+2}{2} \\
D-n+2
\end{array} \bigg|[x_{-}-x_{+}]_{a_1,a_2}\cdot x\right)\\
			&+(\textbf{1})_{m_2}(n;a_1;a_2;m_1)\cdot \ _3F_2\left(\begin{array}{l}
1,\frac{D-n+2}{2},m_1+D-n+1 \\
D-n+2,m_1+D-n+2
\end{array} \bigg|[x_{-}-x_{+}]_{a_1,a_2}\cdot x\right)\Big\}dx,
		\end{aligned}
	\end{equation}
	where the expansion coefficients are
	
	\begin{equation}
		\begin{aligned}
			(\boldsymbol{1})_{m_2}(n;a_1;a_2,m_1)=&\frac{(D-n-1)\cdot\Big(\textbf{P}_1(a_1;a_2)\cdot \textbf{N}^{(1)}_{m_2-1}(a_1;a_2)+\textbf{P}_2(a_1;a_2)\cdot \textbf{N}^{(1)}_{m_2-2}(a_1;a_2)\Big)}{2(D-n+1+m_1)},\\
			(\boldsymbol{0})_{m_2}(n;a_1;a_2;m_1)=&\textbf{Q}_0(a_1;a_2)\cdot \textbf{N}^{(0)}_{m_2}(a_1;a_2)+\textbf{Q}_1(a_1;a_2)\cdot \textbf{N}^{(0)}_{m_2-1}(a_1;a_2).
		\end{aligned}
		\label{eq:define 10-1}
	\end{equation}
	The $\textbf{N}^{(1)}_{m'}(a_1;a_2)$ and $\textbf{N}^{(2)}_{m'}(a_1;a_2)$ come from the Taylor expansion of  terms like $(1+A_1x+A_2x^2)^{c_1}(1+B_1x+B_2x^2)^{c_2}$,which are
 
		\begin{align}
			&(1+\textbf{A}_1(a_1;a_2)x+\textbf{A}_2(a_1;a_2)x^2)^{-\frac{D-n}{2}}(1+\textbf{B}_1(a_1;a_2)x+\textbf{B}_2(a_1;a_2)x^2)^{\frac{D-n-3}{2}}\notag\\
			=&\sum_{m'=0}^\infty \textbf{N}^{(1)}_{m'}(a_1;a_2) \cdot x^{m'},\notag\\
			&(1+\textbf{A}_1(a_1;a_2)x+\textbf{A}_2(a_1;a_2)x^2)^{-\frac{D-n}{2}}(1+\textbf{B}_1(a_1;a_2)x+\textbf{B}_2(a_1;a_2)x^2)^{\frac{D-n-1}{2}}\notag\\
			=&\sum_{m'=0}^\infty \textbf{N}^{(0)}_{m'}(a_1;a_2) \cdot x^{m'}.
		\end{align}
	By formula \eqref{eq:formula 01}, the Taylor expansion coefficients are 

		\begin{align}
			\textbf{N}^{(1)}_{m'}(a_1;a_2)=&\sum_{l=0}^{m'}\sum_{i=0}^{\lfloor\frac{l}{2}\rfloor}\sum_{j=0}^{\lfloor\frac{m'-l}{2}\rfloor}\frac{(-\frac{D-n}{2}-l+i+1)_{(l-i)}(\frac{D-n-3}{2}-m'+l+j+1)_{(m'-l-j)}}{(l-2i)!\cdot i!(m'-l-2j)!\cdot j!}\notag\\
   &\times \textbf{A}_1(a_1;a_2)^{l-2i}\textbf{A}_2(a_1;a_2)^{i}\textbf{B}_1(a_1;a_2)^{m'-l-2j}\textbf{B}_2(a_1;a_2)^{j},\notag\\
			\textbf{N}^{(0)}_{m'}(a_1;a_2)=&\sum_{l=0}^{m'}\sum_{i=0}^{\lfloor\frac{l}{2}\rfloor}\sum_{j=0}^{\lfloor\frac{m'-l}{2}\rfloor}\frac{(-\frac{D-n}{2}-l+i+1)_{(l-i)}(\frac{D-n-1}{2}-m'+l+j+1)_{(m'-l-j)}}{(l-2i)!\cdot i!\cdot(m'-l-2j)!\cdot j!}\notag\\
   &\times \textbf{A}(a_1;a_2)_1^{l-2i}\textbf{A}_2(a_1;a_2)^{i}\textbf{B}_1(a_1;a_2)^{m'-l-2j}\textbf{B}_2(a_1;a_2)^{j}.\label{eq:define N}
		\end{align}
		
	\textbf{Again, we need to emphasize label $a_1,a_2$ are not symmetric} in $\textbf{N}^{(1)}_{m'}(a_1;a_2)$, $\textbf{N}^{(0)}_{m'}(a_1;a_2)$ and $(\boldsymbol{1})_{m_2}(n;a_1;a_2,m_1)$, $(\boldsymbol{0})_{m_2}(n;a_1;a_2,m_1)$.
	By using integral formula\eqref{eq:formula 02}, we can calculate the integral \eqref{eq:n to n-2 01}. Eventually, by summing everything up, we obtain the final generating function of $n$-gon to $(n-2)$-gon as

		\begin{align}
			&\textbf{GF}_{n\to n;\widehat{a_1,a_2}}(\mathcal{t})\notag\\
            =&(1-x_{+}\cdot \mathcal{t})^{\frac{-2+D-n}{2}}(1-x_{-}\cdot \mathcal{t})^{\frac{-2+D-n}{2}}\notag\\
			\times &\sum_{m_1,m_2=0}^{\infty}\Bigg\{\frac{1}{m_1+m_2+D-n+1}
			\cdot\frac{\mathcal{t}^{m_1+m_2+2}}{(1-[x_{+}]_{a_1,a_2}\cdot \mathcal{t})^{m_1+m_2+D-n+1}}\notag\\
			\times &\Big\{\left((\textbf{1})_{m_2}(n;a_1;a_2;m_1)[\textbf{C}_{a_1}^{(1)}(n-1;m_1)]_{a_2}+(\textbf{1})_{m_2}(n;a_2;a_1;m_1)[\textbf{C}_{a_2}^{(1)}(n-1;m_1)]_{a_1}\right)\notag\\
   &\ \ \ \times\ _4F_3\left(\begin{array}{l}
1,\frac{D-n+2}{2},m_1+D-n+1,m_1+m_2+D-n+1 \notag\\
D-n+2,m_1+D-n+2,m_1+m_2+D-n+2
\end{array} \bigg|\frac{[x_{-}-x_{+}]_{a_1,a_2}\cdot\mathcal{t}}{1 - [x_+]_{a_1,a_2}\cdot \mathcal{t}}\right)\notag\\
			&+\left((\textbf{1})_{m_2}(n;a_1;a_2;m_1)[\textbf{C}_{a_1}^{(2)}(n-1;m_1)]_{a_2}+(\textbf{1})_{m_2}(n;a_2;a_1;m_1)[\textbf{C}_{a_2}^{(2)}(n-1;m_1)]_{a_1}\right)\notag\\
   &\ \ \ \times\ _4F_3\left(\begin{array}{l}
2,\frac{D-n+4}{2},m_1+D-n+1,m_1+m_2+D-n+1 \notag\\
D-n+3,m_1+D-n+2,m_1+m_2+D-n+2
\end{array} \bigg|\frac{[x_{-}-x_{+}]_{a_1,a_2}\cdot\mathcal{t}}{1 - [x_+]_{a_1,a_2}\cdot \mathcal{t}}\right)\notag\\
			&+\left((\textbf{0})_{m_2}(n;a_1;a_2;m_1)[\textbf{C}_{a_1}^{(1)}(n-1;m_1)]_{a_2}+(\textbf{0})_{m_2}(n;a_2;a_1;m_1)[\textbf{C}_{a_2}^{(1)}(n-1;m_1)]_{a_1}\right)\notag\\
   &\ \ \ \times\ _3F_2\left(\begin{array}{l}
1,\frac{D-n+2}{2},m_1+m_2+D-n+1\notag \\
D-n+2,m_1+m_2+D-n+2
\end{array} \bigg|\frac{[x_{-}-x_{+}]_{a_1,a_2}\cdot\mathcal{t}}{1 - [x_+]_{a_1,a_2}\cdot \mathcal{t}}\right)\notag\\
			&+\left((\textbf{0})_{m_2}(n;a_1;a_2;m_1)[\textbf{C}_{a_1}^{(2)}(n-1;m_1)]_{a_2}+(\textbf{0})_{m_2}(n;a_2;a_1;m_1)[\textbf{C}_{a_2}^{(2)}(n-1;m_1)]_{a_1}\right)\notag\\
   &\ \ \ \times\ _3F_2\left(\begin{array}{l}
2,\frac{D-n+4}{2},m_1+m_2+D-n+1 \\
D-n+3,m_1+m_2+D-n+2
\end{array} \bigg|\frac{[x_{-}-x_{+}]_{a_1,a_2}\cdot\mathcal{t}}{1 - [x_+]_{a_1,a_2}\cdot \mathcal{t}}\right)\Big\}\Bigg\}.	\label{eq:GF n to n-2}
		\end{align}

	\section{Generating function of $n$-gon to $(n-k)$-gon}
 \label{sec:n to n-k}

From\eqref{eq:GF n to n-1} and\eqref{eq:GF n to n-2}, we found $\textbf{GF}_{n\to n;\widehat{a_1,a_2}}(\mathcal{t})$ has an analogous functional form with $\textbf{GF}_{n\to n;\widehat{a_1}}(\mathcal{t})$. This suggests that the generating function $\textbf{GF}_{n\to n;\widehat{\textbf{I}_k}}(\mathcal{t})$ (where the label list $\textbf{I}_k=\{a_1,a_2,...,a_k\}$) should also have the same functional form. In this section, we will provide the form of $\textbf{GF}_{n\to n;\widehat{\textbf{I}_k}}(\mathcal{t})$ directly. A proof by induction will be given in the next section to show it does satisfy the recursion relation \eqref{eq:relation 01}.

Before that, we provide some notations to represent generalized hypergeometric functions. First, we construct an array including $k$ elements$\{b_1,b_2,...,b_{k}\}$with each $b_i$ is either $0$ or $1$. For example, for $k=4$, the array can be one of 

\begin{equation}
\begin{aligned}
    \{0,0,0,0\},\{0,0,0,1\}, \{0,0,1,0\}, \{0,0,1,1\}, \{0,1,0,0\}, \{0,1,0,1\}, \{0,1,1,0\}, \{0,1,1,1\},\\
    \{1,0,0,0\}, \{1,0,0,1\}, \{1,0,1,0\},  \{1,0,1,1\}, \{1,1,0,0\}, \{1,1,0,1\},  \{1,1,1,0\}, \{1,1,1,1\}.
\end{aligned}
\end{equation}
Now we can construct two types of generalized hypergeometric functions based on the array as follows:
 
	\begin{equation}
		\begin{aligned}
			&\textbf{HG}_1(n,k;\{b_1,\cdots,b_{k}\};z)=_{(2+k)}F_{(1+k)}\left(\begin{array}{l}
1,\frac{D-n+k}{2},S_1-b_1,S_2-b_2,\cdots,S_k-b_k\\
D-n+k,S_1,S_2,\cdots,S_k
\end{array} \bigg|z\right),\\
	&\textbf{HG}_2(n,k;\{b_1,\cdots,b_{k}\};z)
			=_{(2+k)}F_{(1+k)}\left(\begin{array}{l}
2,\frac{D-n+k+2}{2},S_1-b_1,S_2-b_2,\cdots,S_k-b_k\\
D-n+k+1,S_1,S_2,\cdots,S_k
\end{array} \bigg|z\right),
		\end{aligned}    
		\label{eq:define hp}
	\end{equation}
	where 
 \begin{equation}
 S_i=\sum_{j=1}^im_j+D-n+k.
 \end{equation}
For example, 
	
	\begin{equation}
		\begin{aligned}
			&\textbf{HG}_1(n,3;\{0,0,1\};z)\\
			=& _5F_4\left(\begin{array}{l}
1,\frac{D-n+3}{2},m_1+D-n+3,m_1+m_2+D-n+3,m_1+m_2+m_3+D-n+2\\
D-n+3,m_1+D-n+3,m_1+m_2+D-n+3,m_1+m_2+m_3+D-n+3
\end{array} \bigg|z\right)\\
			=&_3F_2\left(\begin{array}{l}
1,\frac{D-n+3}{2},m_1+m_2+m_3+D-n+2\\
D-n+3,m_1+m_2+m_3+D-n+3
\end{array} \bigg|z\right),\\
&\ \\
			&\textbf{HG}_2(n,3;\{1,0,1\};z)\\
			=& _5F_4\left(\begin{array}{l}
2,\frac{D-n+5}{2},m_1+D-n+2,m_1+m_2+D-n+3,m_1+m_2+m_3+D-n+2\\
D-n+4,m_1+D-n+3,m_1+m_2+D-n+3,m_1+m_2+m_3+D-n+3
\end{array} \bigg|z\right)\\
			=&_4F_3\left(\begin{array}{l}
2,\frac{D-n+5}{2},m_1+D-n+2,m_1+m_2+m_3+D-n+2\\
D-n+4,m_1+D-n+3,m_1+m_2+m_3+D-n+3
\end{array} \bigg|z\right).
		\end{aligned}
	\end{equation}
	
	Then generating function $\textbf{GF}_{n\to n;\widehat{\textbf{I}_k}}$ is
		\begin{align}
			&\textbf{GF}_{n\to n;\widehat{\textbf{I}_k}}(\mathcal{t})
         =(1-x_{+}\cdot \mathcal{t})^{\frac{-2+D-n}{2}}\cdot (1-x_{-}\cdot \mathcal{t})^{\frac{-2+D-n}{2}}\notag\\
			&\times \sum_{m_1,...,m_k=0}^{\infty}\Bigg\{\frac{1}{\sum_{i=1}^k m_i+D-n+k-1}\cdot\frac{\mathcal{t}^{\sum_{i=1}^k m_i+k}}{(1-[x_{+}]_{\textbf{I}_k}\cdot \mathcal{t})^{\sum_{i=1}^k m_i+D-n+k-1}}\notag\\	
			&\times \sum_{\{a'_1,...,a'_k\}\in \sigma(\textbf{I}_k),}\ \sum_{b_1,b_2,...,b_{k-1}=0}^1\textbf{C}_{\{b_1,...,b_{k-1}\}}^{(a'_1,...,a'_k)}(n)\notag\\
			&\times \Big\{[\textbf{C}_{a'_1}^{(1)}(n-k+1;m_1)]_{a'_2,...,a'_k} \cdot \textbf{HG}_1(n,k;\{b_1,b_2,...,b_{k-1},1\};\mathcal{W}_{\textbf{I}_k}(\mathcal{t}))\notag\\
			&\ \ \ \ +[\textbf{C}_{a'_1}^{(2)}(n-k+1;m_1)]_{a'_2,...,a'_k}\cdot \textbf{HG}_2(n,k;\{b_1,b_2,...,b_{k-1},1\};\mathcal{W}_{\textbf{I}_k}(\mathcal{t}))\Big\}\Bigg\},\label{eq:GF n to n-k}
		\end{align}

 where \begin{equation}
		\mathcal{W}_{\textbf{I}_k}(\mathcal{t})=\frac{[x_{-}-x_{+}]_{\textbf{I}_k}\cdot\mathcal{t}}{1 - [x_+]_{\textbf{I}_k}\cdot \mathcal{t}}.
	\end{equation}
	
	The first summation in the penultimate line of \eqref{eq:GF n to n-k} runs over all the permutations of $\textbf{I}_k=\{a_1,a_2,...,a_k\}$. For example for $k=3$, this summation should run over 
	
	\begin{equation}
		\begin{aligned}
			\{a'_1,a'_2,a'_3\}=\Big\{\{a_1,a_2,a_3\},\ \{a_1,a_3,a_2\},\ \{a_2,a_1,a_3\},\ \{a_2,a_3,a_1\},\ \{a_3,a_1,a_2\},\ \{a_3,a_2,a_1\}\Big\}.
		\end{aligned}
	\end{equation}
	It's important to note that the second summation on that line does \textbf{not} include $b_k$.
 For example for $k=4$, this summation runs over without $b_4$ but
	
	\begin{equation}
		\{b_1,b_2,b_3\}=\Big\{\{0,0,0\},\{0,0,1\},\{0,1,0\},\{0,1,1\},\{1,0,0\},\{1,0,1\},\{1,1,0\},\{1,1,1\}\Big\}.
	\end{equation}
	For a given permutation $(a'_1,a'_2...,a'_k)$ and a fixed $\{b_1,b_2,...,b_{k-1}\}$, the $\textbf{C}^{(a'_1,...,a'_k)}_{\{b_1,...,b_{k-1}\}}(n)$ in the penultimate line is constructed as 
	
	\begin{equation}
		\begin{aligned}
			&\textbf{C}^{(a'_1,...,a'_k)}_{\{b_1,...,b_{k-1}\}}(n)\\
			=&[(\textbf{b}_1)_{m_2}(n-k+2;1;a'_1;a'_2;m_1)]_{a'_3,...,a'_k}\\
			\cdot &[(\textbf{b}_2)_{m_3}(n-k+3;2;\{a'_1,a'_2\};a'_3;m_1+m_2)]_{a'_4,...,a'_k}\\
			&\cdots\\
			\cdot & [(\textbf{b}_i)_{m_{i+1}}(n-k+i+1;i;\{a'_1,...,a'_i\};a'_{i+1};m_1+\cdots+m_i)]_{a'_{i+2},...,a'_k}\\
			&\cdots\\
			\cdot &(\textbf{b}_{k-1})_{m_k}(n;k-1;\{a'_1,...,a'_{k-1}\};a'_k;m_1+...+m_{k-1}).
		\end{aligned}
		\label{eq:define CCC}
	\end{equation}
	For $k=4$, $\{b_1,b_2,b_3\}=\{0,1,0\}$ as an example, the $\textbf{C}^{(a'_1,...,a'_4)}_{\{0,1,0\}}(n)$ is\footnote{Remember notation $[\Delta]_{\textbf{b}}$ means we add a subscript $\textbf{b}$ on each term in $\Delta$ with form $(\overline{AB})$ or $(\overline{AB})_{\textbf{a}}$. As we explained in second chapter.}
	
	\begin{equation}
		\begin{aligned}
			\textbf{C}^{(a'_1,...,a'_4)}_{\{0,1,0\}}(n)=&[(\textbf{0})_{m_2}(n-2;1;a'_1;a'_2;m_1)]_{a'_3,a'_4}\cdot [(\textbf{1})_{m_3}(n-1;2;\{a'_1,a'_2\};a'_3;m_1+m_2)]_{a'_4}\\
			\cdot &(\textbf{0})_{m_4}(n;3;\{a'_1,a'_2,a'_3\};a'_4;m_1+m_2+m_3).
		\end{aligned}
	\end{equation}
	There is a slight difference between the definitions of $\textbf{0}$ and $\textbf{1}$ when compared to the definitions in formal section \eqref{eq:define 10-1}. Here, it includes one more variable $k'$ , defined as
	
	\begin{equation}
		\begin{aligned}
			(\boldsymbol{1})_{m'}(n';k';\textbf{a};b;sum)=&\frac{(D-n'-1)\cdot\left(\textbf{P}_1(\textbf{a};b)\cdot \textbf{N}^{(1)}_{m'-1}(\textbf{a};b)+\textbf{P}_2(\textbf{a};b)\cdot \textbf{N}^{(1)}_{m'-2}(\textbf{a};b)\right)}{2(D-n'+k'+sum)},\\
			(\boldsymbol{0})_{m'}(n';k';\textbf{a};b;sum)=&\textbf{Q}_0(\textbf{a};b)\cdot \textbf{N}^{(0)}_{m'}(\textbf{a};b)+\textbf{Q}_1(\textbf{a};b)\cdot \textbf{N}^{(0)}_{m'-1}(\textbf{a};b).
		\end{aligned}
		\label{eq:define 10-2}
	\end{equation}

The variables $n'$, $k'$, $m'$, and $sum$ must be non-negative integers. The variable $\mathbf{a}$ can be fixed either a single label or a list of labels, while $b$ solely represents a single label. Upon setting $m'=m_2$, $n'=n$, $k=1$, $\mathbf{a}=a_1$, $b=a_2$, and $sum=m_1$, we revert to the form \eqref{eq:define 10-1} as outlined in the previous section. The other coefficients $\mathbf{P}_1(\mathbf{a};b)$, $\mathbf{P}_2(\mathbf{a};b)$, $\mathbf{Q}_0(\mathbf{a};b)$, $\mathbf{Q}_1(\mathbf{a};b)$, $\mathbf{N}_{m'}(\mathbf{a};b)$ are defined by merely replacing all instances of the labels $a_1$ and $a_2$ in \eqref{eq:define ABPQ}, and \eqref{eq:define N} with variables $\mathbf{a}$ and  $b$ respectively. We have thus completed the explanation of every term in $\mathbf{GF}_{n\to n;\widehat{\mathbf{I}_k}}(\mathcal{t})$ \eqref{eq:GF n to n-k}. Furthermore, if we additionally define $\mathbf{C}_{\emptyset}^{a_1}\equiv 1$ for $k=1$, we find that the function \eqref{eq:GF n to n-k} aligns with $\mathbf{GF}_{n\to n;\widehat{a_1}}(\mathcal{t})$\eqref{eq:GF n to n-1} and $\mathbf{GF}_{n\to n;\widehat{a_1,a_2}}(\mathcal{t})$\eqref{eq:GF n to n-2}.

\subsection{Analytic example}
\subsubsection{Triangle to Tadpole}
Now we present the reduction of an tensor triangle to tadpole $\textbf{GF}_{3 \to 3:\widehat{2,3}}$ as an example to illustrate the analyticity. The generating function can be obtain from \eqref{eq:GF n to n-k} by selecting $n=3,k=2,a_1=2,a_2=3$. Expanding the generating function as a series of $\mathcal{t}$, we can obtain:

\begin{equation}
    \textbf{GF}_{3\to 3;\widehat{2,3}}(\mathcal{t})=\sum_{r=0}^{\infty} \mathcal{t}^r\cdot C^{(r)}_{3 \to 3;\widehat{2,3}},
\end{equation}
where $C^{(r)}_{3\to 3;\widehat{2,3}}$ are the reduction coefficients of tensor triangle

\begin{equation}
    I^{(r)}_3=\int\frac{d^Dl}{i\pi^{D/2}}\frac{(2R\cdot l)^r}{((l-q_1)^2-M_1^2)((l-q_2)^2-M_2^2)((l-q_3)^2-M_3^2)},
\end{equation}
to the master integral

\begin{equation}
   I^{(0)}_1=\int\frac{d^Dl}{i\pi^{D/2}}\frac{1}{(l-q_1)^2-M_1^2}.
\end{equation}
Moreover, we list the first two non-zero orders in the expansion:

\begin{equation}
\begin{aligned}
      C^{(2)}_{3\to 3;\widehat{2,3}}=&X^{(2)}\cdot[X^{(3)}]_{2}+X^{(3)}\cdot [X^{(2)}]_{3},\\
      C^{(3)}_{3\to 3;\widehat{2,3}}=&\frac{1}{2(D-1)}\Bigg\{D\left(X^{(2)}\cdot[X^{(3)}]_2\cdot[x_{+}+x_{-}]_2+X^{(3)}\cdot[X^{(2)}]_3\cdot[x_{+}+x_{-}]_3\right)\\+&\left(X^{(2)}\cdot[X^{(3)}]_{2}+X^{(3)}\cdot [X^{(2)}]_{3}\right)\big((D-1)\cdot [x_{+}+x_{-}]_{2,3}+(D+1)\cdot (x_{+}+x_{-})\big)\\
      +&8\left(Y^{(2)}\cdot[X^{(3)}]_{2}+Y^{(3)}\cdot [X^{(2)}]_{3}\right)+4\left(X^{(2)}\cdot[Y^{(3)}]_{2}+X^{(3)}\cdot [Y^{(2)}]_{3}\right)\Bigg\}.\\
\end{aligned}
\end{equation}
\subsubsection{Reduction Coefficients of  $n$-gon to $(n-k)$-gon}
In addition, for the general reduction of an $n$-gon to an $(n-k)$-gon for $k\geq 1$, we expand

\begin{equation}
    \textbf{GF}_{n\to n;\widehat{\textbf{I}_k}}(\mathcal{t})=\sum_{r=0}^\infty \mathcal{t}^r\cdot C^{(r)}_{n\to n;\widehat{\textbf{I}_k}}.
\end{equation}
The reduction coefficients are
\begin{equation}
\begin{aligned}
  C^{(r)}_{n\to n;\widehat{\textbf{I}_k}}=&\sum_{m_1,...,m_k=0}^{\sum_{i=1}^km_i+k\leq r}\Bigg\{\sum_{l_1+l_2+l_3+\sum_{i=1}^km_i+k=r}\textbf{N}_1^{(l_1)}\Bigg(\textbf{N}_2^{(l_2)}(m_1,...,m_k)\\
    \times &\sum_{\{a'_1,...,a'_k\}\in \sigma(\textbf{I}_k),}\sum_{b_1,b_2,...,b_{k-1}=0}^1\textbf{C}_{\{b_1,...,b_{k-1}\}}^{(a'_1,...,a'_k)}(n)\\
   \times&\Big([\textbf{C}_{a'_1}^{(1)}(n-k+1;m_1)]_{a'_2,...,a'_k} \cdot \textbf{M}^{(l_3)}_{1;\{b_1,b_2,...,b_{k-1}\}}(m_1,...,m_k)\\
   &\ +[\textbf{C}_{a'_1}^{(2)}(n-k+1;m_1)]_{a'_2,...,a'_k}\cdot \textbf{M}^{(l_3)}_{2;\{b_1,b_2,...,b_{k-1}\}}(m_1,...,m_k)\Big)\Bigg)\Bigg\}.
\end{aligned}
\end{equation}
The sources of those coefficients are shown as follows:
\begin{itemize}
    \item $\textbf{N}_1^{(l_1)}$ comes from
    \begin{equation}
    \begin{aligned}
        &(1-x_{+}\cdot \mathcal{t})^{\frac{-2+D-n}{2}}(1-x_{-}\cdot \mathcal{t})^{\frac{-2+D-n}{2}}\\
        =&\sum_{l_1=0}^\infty\Bigg(\sum_{i=0}^{l_1}\frac{(-x_{+})^{i}(\frac{-2+D-n}{2}-i+1)_{i}}{i!}\cdot \frac{(-x_{-})^{l_1-i}(\frac{-2+D-n}{2}-(l_1-i)+1)_{(l_1-i)}}{(l_1-i)!}\Bigg)\mathcal{t}^{l_1}\\
        =&\sum_{l_1=0}^\infty\textbf{N}^{(l_1)}_1\cdot \mathcal{t}^{l_1}.
    \end{aligned}
    \end{equation}

\item $\textbf{N}_2^{(l_2)}(m_1,...,m_k)$ comes from
\begin{equation}
\begin{aligned}
&\frac{1}{\sum_{i=1}^k m_i+D-n+k-1}\frac{\mathcal{t}^{\sum_{i=1}^k m_i+k}}{(1-[x_{+}]_{\textbf{I}_k}\cdot \mathcal{t})^{\sum_{i=1}^k m_i+D-n+k-1}}\\
=&\sum _{l_2=0}^\infty\left(\frac{([x_{+}]_{\textbf{I}_k})^{l_2}\left(\sum_{i=1}^k m_i+D-n+k-1\right)_{l_2}}{\left(\sum_{i=1}^k m_i+D-n+k-1\right)\cdot l_2!}\right)\cdot \mathcal{t}^{l_2+\sum_{i=1}^k m_i+k}\\
=&\sum_{l_2=0}^\infty\textbf{N}_2^{(l_2)}(m_1,...,m_k)\cdot \mathcal{t}^{l_2+\sum_{i=1}^k m_i+k}.
\end{aligned}
\end{equation}

\item By \eqref{eq:Taylor 01}, $\textbf{M}^{(l_3)}_{1;\{b_1,b_2,...,b_{k-1}\}}(m_1,...,m_k)$ and $\textbf{M}^{(l_3)}_{2;\{b_1,b_2,...,b_{k-1}\}}(m_1,...,m_k)$ come from

\begin{align}
    &\textbf{HG}_1(n,k;\{b_1,b_2,...,b_{k-1},1\};\mathcal{W}_{\textbf{I}_k}(\mathcal{t}))\notag \\
    =& 1+\sum_{l_3=1}^{\infty}\Bigg(\sum_{i=1}^{l_3}\frac{\textbf{C}^{i-1}_{l_3-1}(\frac{D-n+k}{2})_i([x_{-}-x_{+}]_{\textbf{I}_k})^i([x_{+}]_{\textbf{I}_k})^{l_3-i}}{(D-n+k)_i}\cdot \mathcal{S}_{i}(\{b_1,...,b_{k-1}\})\Bigg)\mathcal{t}^{l_3}\notag\\ =&\sum_{l_3=0}^\infty\left(\textbf{M}^{(l_3)}_{1;\{b_1,b_2,...,b_{k-1}\}}(m_1,...,m_k)\right)\cdot \mathcal{t}^{l_3},
\end{align}
\begin{align}
    &\textbf{HG}_2(n,k;\{b_1,b_2,...,b_{k-1},1\};\mathcal{W}_{\textbf{I}_k}(\mathcal{t}))\notag \\
    =& 1+\sum_{l_3=1}^{\infty}\Bigg(\sum_{i=1}^{l_3}\frac{\textbf{C}^{i-1}_{l_3-1}(\frac{D-n+k+2}{2})_i([x_{-}-x_{+}]_{\textbf{I}_k})^i([x_{+}]_{\textbf{I}_k})^{l_3-i}(i+1)}{(D-n+k+1)_i}\cdot \mathcal{S}_{i}(\{b_1,...,b_{k-1}\})\Bigg)\mathcal{t}^{l_3}\notag\\ =&\sum_{l_3=0}^\infty\left(\textbf{M}^{(l_3)}_{2;\{b_1,b_2,...,b_{k-1}\}}(m_1,...,m_k)\right)\cdot \mathcal{t}^{l_3},
\end{align}
where the factor $\mathcal{S}_{i}(\{b_1,...,b_{k-1}\})$ are defined as
\begin{equation}
\begin{aligned}
    &\mathcal{S}_{i}(\{b_1,...,b_{k-1}\})\\
    =&\frac{\left(\prod_{\alpha=1}^{k-1}(\sum_{\beta=1}^\alpha m_\beta+D-n+k-b_{\alpha})_i\right)\left(\sum_{\beta=1}^k m_\beta+D-n+k-1\right)_i}{\left(\prod_{\alpha=1}^{k-1}(\sum_{\beta=1}^\alpha m_\beta+D-n+k)_i\right)\left(\sum_{\beta=1}^k m_\beta+D-n+k\right)_i}.
\end{aligned}
\end{equation}

\end{itemize}

	\section{Proof}
 \label{sec:proof}

In this section, we will provide an inductive proof of parameter $k$ to verify the correctness of \eqref{eq:GF n to n-k}. The main methodology echoes the computation of $\mathbf{GF}_{n\to n;\widehat{a_1,a_2}}(\mathcal{t})$. Suppose that expression \eqref{eq:GF n to n-k} holds for $k$. We can evaluate the generating function for reduction of an $n$-gon to $(n-(k+1))$-gon by solving the differential equation \eqref{eq:relation 02}. After writing down the general solution, we divided the integration into several parts according to the generalized hypergeometric function. During the steps of integration,  we firstly alter the integration variable.   Subsequently, we conduct the Taylor expansion on a part of the integrand function. Lastly, by selecting the suitable integration formula\eqref{eq:formula 02}, we affirm that the statement \eqref{eq:GF n to n-k} holds true for $k+1$. Readers who are already familiar with this process may choose to skip this section. The derivation and integration rules that we employ during our proof are outlined in \eqref{eq:dev rule 01} and \eqref{eq:formula 02}. In the context of \eqref{eq:define hp}, these two rules transform into:

	\begin{equation}
		\begin{aligned}
			&\frac{d}{dx}\textbf{HG}_{l}(n-1;k;\{b_1,...,b_{k-1},1\};z)\\
			=&\frac{\sum_{j=1}^k m_j+D-n-k}{z}\Big(\textbf{HG}_{l}(n-1;k;\{b_1,...,b_{k-1},0\};z)-\textbf{HG}_{l}(n-1;k;\{b_1,...,b_{k-1},1\};z)\Big),\\
            &\text{for}\ l=1,2,
		\end{aligned}
        \label{eq:dev formula 03}
	\end{equation}
	and
	
	\begin{equation}
		\begin{aligned}
			&\int dz\ z^{\sum_{j=1}^{k+1}m_j+D-n-k-1}\cdot\textbf{HG}_{l}(n-1;k;\{b_1,...,b_{k-1},l\};h\cdot z)\\
			=&\frac{z^{\sum_{j=1}^{k+1}m_j+D-n-k}}{\sum_{j=1}^{k+1}m_j+D-n-k}\cdot\textbf{HG}_{l}(n;k+1;\{b_1,...,b_{k-1},l,1\}; h\cdot z),\ \text{for}\ l=1,2.
		\end{aligned}
        \label{eq:int formula 04}
	\end{equation}
	
	Supposing generating function of $n$-gon to $(n-k)$-gon  $\textbf{GF}_{n\to n;\textbf{I}_k}(\mathcal{t})$ satisfies \eqref{eq:GF n to n-k}, then recursive relation of $\textbf{GF}_{n\to n;\widehat{\textbf{I}_{k+1}}}(\mathcal{t})$ with $\textbf{I}_{k+1}=\{a_1,a_2,...,a_{k+1}\}$ is
	
	\begin{equation}
		\begin{aligned}
			&\left((D-n-1)-2(D-n)\cdot\frac{(\overline{VL})}{(\overline{LL})}\mathcal{t}-4\cdot\frac{R^2-(\overline{VV})}{(\overline{LL})}\mathcal{t}^2\right)\textbf{GF}_{n \to n;\widehat{\textbf{I}_{k+1}}}(\mathcal{t})\\
			+&\left(\mathcal{t}-4\cdot\frac{(\overline{VL})}{(\overline{LL})}\mathcal{t}^2-4\cdot\frac{R^2-(\overline{VV})}{(\overline{LL})}\mathcal{t}^3\right)\textbf{GF}'_{n\to n;\widehat{\textbf{I}_{k+1}}}(\mathcal{t})\\
			=&\sum_{a_i,i=1}^{k+1}\Bigg\{2Y^{(a_i)}\left(\mathcal{t}^3\cdot\textbf{GF}'_{n;\widehat{a_i}\to n;\widehat{\textbf{I}_{k+1}}}(\mathcal{t})+\mathcal{t}^2\cdot\textbf{GF}_{n;\widehat{a_i}\to n;\widehat{\textbf{I}_{k+1}}}(\mathcal{t})\right)\\
			+&X^{(a_i)}\left((D-n)\mathcal{t}\cdot\textbf{GF}_{n;\widehat{a_i}\to n;\widehat{\textbf{I}_{k+1}}}(\mathcal{t})+\mathcal{t}^2\cdot\textbf{GF}'_{n;\widehat{a_i}\to n;\widehat{\textbf{I}_{k+1}}}(\mathcal{t})\right)\Bigg\}.
		\end{aligned}
		\label{eq:dif eq n to n-k-1}
	\end{equation}
	The general solution is
	\begin{equation}
		\begin{aligned}
			&\mathcal{t}^{1-D+n}(1-x_{+}\cdot \mathcal{t})^{\frac{-2+D-n}{2}}(1-x_{-}\cdot \mathcal{t})^{\frac{-2+D-n}{2}}\\
			\times & \Bigg\{C_1+\int_0^\mathcal{t}\mathcal{u}^{-1+D-n}(1-x_{+}\cdot \mathcal{u})^{\frac{n-D}{2}}(1-x_{-}\cdot \mathcal{u})^{\frac{n-D}{2}}\\
			\times&\sum_{a_i,i=1}^{k+1}\Big\{2Y^{(a_i)}\left(\mathcal{u}^2\cdot\textbf{GF}'_{n;\widehat{a_i}\to n;\widehat{\textbf{I}_{k+1}}}(\mathcal{u})+\mathcal{u}\cdot\textbf{GF}_{n;\widehat{a_i}\to n;\widehat{\textbf{I}_{k+1}}}(\mathcal{u})\right)\\
			+&X^{(a_i)}\left((D-n)\cdot\textbf{GF}_{n;\widehat{a_i}\to n;\widehat{\textbf{I}_{k+1}}}(\mathcal{u})+\mathcal{u}\cdot\textbf{GF}'_{n;\widehat{a_i}\to n;\widehat{\textbf{I}_{k+1}}}(\mathcal{u})\right)\Big\}d\mathcal{u}\Bigg\}.
		\end{aligned}
		\label{eq:GS n to n-k-1}
	\end{equation}
	The undetermined constant $C_1$ is determined to be zero, as the generating function must conform to the requirement of being a Taylor series in terms of $\mathcal{t}$.
Where $\textbf{GF}_{n;\widehat{a_i}\to n;\widehat{\textbf{I}_{k+1}}}(\mathcal{t})=[\textbf{GF}_{n-1\to n-1;\widehat{\textbf{I}_{k+1}/a_i}}(\mathcal{t})]_{a_i}$. We split it into

\begin{equation}
\begin{aligned}
    \textbf{GF}_{n;\widehat{a_i}\to n;\widehat{\textbf{I}_{k+1}}}(\mathcal{t})=&\sum_{m_1,...,m_k=0,}^\infty\sum_{\{a'_1,...,a'_k\}\in \sigma(\textbf{I}_{k+1}/a_i),}\sum_{b_1,...,b_{k-1}=0}^1 [\textbf{C}_{\{b_1,...,b_{k-1}\}}^{(a'_1,...,a'_k)}(n-1)]_{a_i}\\
    \times&
			\left(\sum_{l=1,2}\ [\textbf{C}_{a'_1}^{(l)}(n-k;m_1)]_{a'_2,...,a'_k,a_i}\cdot \textbf{GF}_{n;\widehat{a_i}\to n;\widehat{\textbf{I}_{k+1}}}^{(l);m_1,...,m_k;b_1,...,b_{k-1}}(\mathcal{t})\right),
\end{aligned}
\label{eq:split 01}
\end{equation}
where
 
 \begin{equation}
		\begin{aligned}
			&\textbf{GF}_{n;\widehat{a_i}\to n;\widehat{\textbf{I}_{k+1}}}^{(l);m_1,...,m_k;b_1,...,b_{k-1}}(\mathcal{t})=(1-[x_{+}]_{a_i}\cdot \mathcal{t})^{\frac{-1+D-n}{2}}(1-[x_{-}]_{a_i}\cdot \mathcal{t})^{\frac{-1+D-n}{2}}\\
			&\times \Bigg\{\frac{1}{\sum_{i=1}^k m_i+D-n+k}\cdot \frac{\mathcal{t}^{\sum_{i=1}^k m_i+k}}{(1-[x_{+}]_{\textbf{I}_{k+1}}\cdot \mathcal{t})^{\sum_{i=1}^k m_i+D-n+k}}\\
            &\times\textbf{HG}_{l}(n-1,k;\{b_1,b_2,...,b_{k-1},1\};\mathcal{W}_{\textbf{I}_{k+1}}(\mathcal{t}))\Bigg\},\ \text{for}\ l=1,2.
		\end{aligned}
		\label{eq:GF n to n-k 01}
	\end{equation}
  Since the coefficients $[\textbf{C}^{(a'_1,...,a'_k)}_{\{b_1,...,b_{k-1}\}}(n-1)]_{a_i}$ and $[C^{(1)/(2)}_{a'_1}(n-k;m_1)]_{a'_2,...,a'_k,a_i}$ in \eqref{eq:split 01} are independent of $\mathcal{t}$, they can be pulled out of the integral. Then solution \eqref{eq:GS n to n-k-1} can be separated into

\begin{equation}
\begin{aligned}
&(1-x_{+}\cdot \mathcal{t})^{\frac{-2+D-n}{2}}(1-x_{-}\cdot \mathcal{t})^{\frac{-2+D-n}{2}}\\
\times &\sum_{l=1}^2\sum_{a_i,i=1}^{k+1}\sum_{m_1,...,m_k=0}^\infty\sum_{\{a'_1,...,a'_k\}\in \sigma(\textbf{I}_{k+1}/a_i)}\sum_{b_1,...,b_{k-1}=0}^1 [\textbf{C}_{\{b_1,...,b_{k-1}\}}^{(a'_1,...,a'_k)}(n-1)]_{a_i}[\textbf{C}^{(l)}_{a'_1}(n-k;m_1)]_{a'_2,...,a'_k,a_i}\\
\times&\Bigg\{\mathcal{t}^{1-D+n} \int_0^\mathcal{t}\mathcal{u}^{-1+D-n}(1-x_{+}\cdot \mathcal{u})^{\frac{n-D}{2}}(1-x_{-}\cdot \mathcal{u})^{\frac{n-D}{2}}\\
&\times\Big\{2Y^{(a_i)} \left(\mathcal{u}^2\cdot\textbf{GF}'^{(l);m_1,...,m_k;b_1,...,b_{k-1}}_{n;\widehat{a_i}\to n;\widehat{\textbf{I}_{k+1}}}(\mathcal{u})+\mathcal{u}\cdot\textbf{GF}_{n;\widehat{a_i}\to n;\widehat{\textbf{I}_{k+1}}}^{(l);m_1,...,m_k;b_1,...,b_{k-1}}(\mathcal{u})\right)\\
&+X^{(a_i)} \left((D-n)\cdot\textbf{GF}_{n;\widehat{a_i}\to n;\widehat{\textbf{I}_{k+1}}}^{(l);m_1,...,m_k;b_1,...,b_{k-1}}(\mathcal{u})+\mathcal{u}\cdot\textbf{GF}'^{(l);m_1,...,m_k;b_1,...,b_{k-1}}_{n;\widehat{a_i}\to n;\widehat{\textbf{I}_{k+1}}}(\mathcal{u})\right)\Big\}d\mathcal{u}\Bigg\}.
\end{aligned}
\label{eq:GS n to n-k-1 01}
\end{equation}

Next we focus on the part inside the curly braces. After changing the integration variable $\mathcal{u}$ to $\mathcal{x}_{\textbf{I}_{k+1}}(\mathcal{u})$ :

\begin{equation}
		\mathcal{x}_{\textbf{I}_{k+1}}(\mathcal{u})=\frac{\mathcal{u}}{1 - [x_+]_{\textbf{I}_{k+1}}\cdot \mathcal{u}},
	\end{equation}
and applying derivative formula \eqref{eq:dev formula 03}, terms inside the big curly braces of \eqref{eq:GS n to n-k-1 01} become

	\begin{equation}
		\begin{aligned}
			&\mathcal{t}^{1-D+n}
    \int_0^{\mathcal{x}_{\textbf{I}_{k+1}}(\mathcal{t})}dx\  x^{\sum_{j=1}^k m_j+D-n+k-1}\\
         \times&\Bigg\{\Big\{\left(\textbf{Q}_0(\{a'_1,...,a'_k\};a_i)+\textbf{Q}_1(\{a'_1,...,a'_k\};a_i)\cdot x\right)\\
         &\times \left(1+\textbf{A}_1(\{a'_1,...,a'_k\};a_i)x+\textbf{A}_2(\{a'_1,...,a'_k\};a_i)x^2\right)^{-\frac{D-n}{2}}\\
         &\times \left(1+\textbf{B}_1(\{a'_1,...,a'_k\};a_i)x+\textbf{B}_2(\{a'_1,...,a'_k\};a_i)x^2\right)^{\frac{D-n-1}{2}}\\
         &\times  \textbf{HG}_{l}(n-1;k;\{b_1,b_2,...,b_{k-1},0\};[x_{-}-x_{+}]_{\textbf{I}_{k+1}}\cdot x)\Big\}\\
			+&\Big\{\frac{(D-n-1)\big(\textbf{P}_1(\{a'_1,...,a'_k\};a_i)x+\textbf{P}_2 (\{a'_1,...,a'_k\};a_i)x^2\big)}{2(D-n+k+m_1+m_2+\cdots+m_k)}\\
   &\times \left(1+\textbf{A}_1(\{a'_1,...,a'_k\};a_i)x+\textbf{A}_2(\{a'_1,...,a'_k\};a_i)x^2\right)^{-\frac{D-n}{2}}\\
         &\times \left(1+\textbf{B}_1(\{a'_1,...,a'_k\};a_i)x+\textbf{B}_2(\{a'_1,...,a'_k\};a_i)x^2\right)^{\frac{D-n-3}{2}}\\
   &\times \textbf{HG}_{l}(n-1;k;\{b_1,b_2,...,b_{k-1},1\};[x_{-}-x_{+}]_{\textbf{I}_{k+1}}\cdot x)\Big\}\Bigg\}.
		\end{aligned}
	\end{equation}
The coefficients $\textbf{P}_1(\textbf{a};b)$, $\textbf{P}_2(\textbf{a};b)$, $\textbf{Q}_0(\textbf{a};b)$, $\textbf{Q}_1(\textbf{a};b)$, $\textbf{A}_1(\textbf{a};b)$, $\textbf{A}_2(\textbf{a};b)$, $\textbf{B}_1(\textbf{a};b)$, and $\textbf{B}_2(\textbf{a};b)$ are defined by simply replacing all instances of labels $a_1$ and $a_2$ in  \eqref{eq:define ABPQ}, \eqref{eq:define N} with the label list $\{a'_1,...,a'_k\}$ and the single label $a_i$, respectively.
Then, upon carrying out the series expansion, it transforms into

\begin{equation}
\begin{aligned}
&
    \mathcal{t}^{1-D+n}
   \sum_{m_{k+1}=0}^\infty\ \int_0^{\mathcal{x}_{\textbf{I}_{k+1}}(\mathcal{t})}dx\  x^{\sum_{j=1}^{k+1}m_{j}+D-n+k-1}\\
        \times&\Big\{(\textbf{0})_{m_{k+1}}(n;k;\{a'_1,...,a'_k\};a_i;\sum_{j=1}^k m_j)\cdot \textbf{HG}_{l}(n-1;k;\{b_1,b_2,...,b_{k-1},0\};[x_{-}-x_{+}]_{\textbf{I}_{k+1}}\cdot x)\\
			&+(\textbf{1})_{m_{k+1}}(n;k;\{a'_1,...,a'_k\};a_i;\sum_{j=1}^k m_j)\cdot \textbf{HG}_{l}(n-1;k;\{b_1,b_2,...,b_{k-1},1\};[x_{-}-x_{+}]_{\textbf{I}_{k+1}}\cdot x)\Big\}.
		\end{aligned}
	\end{equation}

According to integration rule \eqref{eq:int formula 04}, above equation becomes
	\begin{equation}
		\begin{aligned}
		&\sum_{m_{k+1}=0}^{\infty}\Bigg\{\frac{1}{\sum_{j=1}^{k+1} m_j+D-n+k}\cdot\frac{\mathcal{t}^{\sum_{j=1}^{k+1} m_j+k+1}}{(1-[x_{+}]_{\textbf{I}_{k+1}}\cdot \mathcal{t})^{\sum_{j=1}^{k+1} m_j+D-n+k}}\\
			\times &\sum_{b_k=0}^1(\textbf{b}_k)_{m_{k+1}}(n;k;\{a'_1,...a'_k\};a_i;\sum_{j=1}^k m_j)\cdot \textbf{HG}_l(n;k+1;\{b_1,b_2,...,b_{k-1},b_k,1\};\mathcal{W}_{\textbf{I}_{k+1}}(\mathcal{t}))\Bigg\}.
		\end{aligned}
		\label{eq:GS n to n-k-1 02}
	\end{equation}
	
When everything is added together, we obtain the generating function of an $n$-gon to an $(n-(k+1))$-gon as follows:

\begin{equation}
    \begin{aligned}
        &\textbf{GF}_{n\to n;\widehat{\textbf{I}_{k+1}}}(\mathcal{t})\\
         =&(1-x_{+}\cdot \mathcal{t})^{\frac{-2+D-n}{2}}(1-x_{-}\cdot \mathcal{t})^{\frac{-2+D-n}{2}}\times\sum_{l=1}^2\sum_{m_1,...,m_{k+1}=0}^\infty\sum_{b_1,...,b_k=0}^1\sum_{a_i,i=1}^{k+1}\sum_{\{a'_1,...,a'_k\}\in \sigma(\textbf{I}_{k+1}/a_i)}\\
         \times &\Bigg\{\frac{1}{\sum_{j=1}^{k+1} m_j+D-n+k}\cdot\frac{\mathcal{t}^{\sum_{j=1}^{k+1} m_j+k+1}}{(1-[x_{+}]_{\textbf{I}_{k+1}}\cdot \mathcal{t})^{{\sum_{j=1}^{k+1} m_j+D-n+k}}}\\
			\times &(\textbf{b}_k)_{m_{k+1}}(n;k;\{a'_1,...a'_k\};a_i;\sum_{j=1}^k m_j)\cdot [\textbf{C}_{\{b_1,...,b_{k-1}\}}^{(a'_1,...,a'_k)}(n-1)]_{a_i}[\textbf{C}^{(l)}_{a'_1}(n-(k+1)-1;m_1)]_{a'_2,...,a'_k,a_i}\\
          \times&\textbf{HG}_l(n;k+1;\{b_1,b_2,...,b_{k-1},b_k,1\};\mathcal{W}_{\textbf{I}_{k+1}}(\mathcal{t}))\Bigg\}.
    \end{aligned}
    \label{eq:GF n to n-k-1 02}
\end{equation}
	
From the definition of $\textbf{C}^{(a'_1,...,a'_k)}_{\{b_1,...,b_{k-1}\}}(n)$ \eqref{eq:define CCC}, we have 

\begin{equation}
    \begin{aligned}
        &[\textbf{C}_{\{b_1,...,b_{k-1}\}}^{(a'_1,...,a'_k)}(n-1)]_{a_i}\cdot(\textbf{b}_k)_{m_{k+1}}(n;k;\{a'_1,...a'_k\};a_i;\sum_{j=1}^k m_j)\\
        =&[(\textbf{b}_1)_{m_2}(n-k+1;1;a'_1;a'_2;m_1)]_{a'_3,...,a'_k,a_i}\\
			\cdot &[(\textbf{b}_2)_{m_3}(n-k+2;2;\{a'_1,a'_2\};a'_3;m_1+m_2)]_{a'_4,...,a'_k,a_i}\\
			&\cdots\\
			\cdot & [(\textbf{b}_j)_{m_{j+1}}(n-k+j;j;\{a'_1,...,a'_j\};a'_{j+1};m_1+\cdots+m_j)]_{a'_{j+2},...,a'_k,a_i}\\
			&\cdots\\
			\cdot &[(\textbf{b}_{k-1})_{m_k}(n;k-1;\{a'_1,...,a'_{k-1}\};a'_k;m_1+...+m_{k-1})]_{a_i}\\
            \cdot &(\textbf{b}_k)_{m_{k+1}}(n;k;\{a'_1,...a'_k\};a_i;\sum_{j=1}^k m_j)\\
            =&\textbf{C}^{(a'_1,...,a'_k,a_i)}_{\{b_1,...,b_{k}\}}(n).
    \end{aligned}
\end{equation}
And obviously, 

\begin{equation}
 \sum_{a_i,i=1}^{k+1}\sum_{\{a'_1,...,a'_k\}\in \sigma(\textbf{I}_{k+1}/a_i)} \textbf{C}^{(a'_1,...,a'_k,a_i)}_{\{b_1,...,b_{k}\}}(n)=\sum_{\{a'_1,...,a'_{k+1}\}\in \sigma(\textbf{I}_{k+1})} \textbf{C}^{(a'_1,...,a'_k,a'_{k+1})}_{\{b_1,...,b_{k}\}}(n).
\end{equation}

Then, we can see that \eqref{eq:GF n to n-k-1 02} precisely aligns with our result \eqref{eq:GF n to n-k} in the $k+1$ case. Therefore, we have successfully validated the correctness of our result.

\section{Conclusion}
	
	\label{sec:con}

In this paper, we present an explicit expression for the generating function for the reduction of an $n$-gon to an $(n-k)$-gon, where $k$ is a general value. We formulate a novel recursive relation of generating functions, which is based on our investigation into Feynman Parametrization in projective space. This newly established relation comprises a single ordinary differential equation in terms of variable $\mathcal{t}$. To solve this equation, it is required to carry out the integral part by the following steps: (1) Changing the integration variable. (2) Implementing the Taylor expansion on the appropriate part of the integrand. (3) Selecting the suitable integration formula. In the end, we unearthed the rule of the general term formula of generating functions.

In addition, there are several comments that we need to supplement. Firstly, it is understood that the reduction coefficients should be \textbf{rational} functions of some Lorentz invariants such as $q_i\cdot q_j, M^2_i$ and spacetime dimension $D$. However, in the expression \eqref{eq:n to n} and \eqref{eq:GF n to n-k}, there exists irrational terms such as the square root term 
\begin{equation}
    [x_{+}-x_{-}]_{\textbf{I}_{k}}=4\sqrt{(\overline{LL})_{\textbf{I}_k}R^2+(\overline{VL})^2_{\textbf{I}_k}-(\overline{LL})_{\textbf{I}_k}(\overline{VV})_{\textbf{I}_k}}/(\overline{LL})_{\textbf{I}_k}.
    \label{eq:irr term}
\end{equation} 
In fact, by the following quadratic transformation

	\begin{equation}
		_2F_1\left(\begin{array}{l}
a , b  \\
 2b 
\end{array} \bigg| z \right)=\left(1-\frac{z}{2}\right)^{-a}\cdot\  _2F_1\left(\begin{array}{l}
\frac{a}{2} , \frac{a+1}{2}  \\
 b+\frac{1}{2}
\end{array} \bigg| \left(\frac{z}{2-z}\right)^2 \right),
		\label{eq:quadratic trans}
	\end{equation}
equation \eqref{eq:n to n} can be changed to a rational form

\begin{equation}
   \mathbf{GF}_{n\to n}(\mathcal{t})=\frac{2}{(x_++x_-)\cdot \mathcal{t}-2} \cdot \ _2F_1\left(\begin{array}{l}
\frac{1}{2} , 1  \\
 \frac{D-n+1}{2}
\end{array} \bigg| \left(\frac{(x_+-x_-)^2\cdot \mathcal{t}^2}{((x_++x_-)\cdot \mathcal{t}-2)^2}\right) \right).
\label{eq:ration n to n}
\end{equation}
We can observe that all terms in the above expression are rational terms. Therefore, even though the generating function we obtain contains irrational terms, it does not alter the fact that the final reduced coefficients remain rational.\footnote{However, for the generating function from an $n$-gon to an $(n-k)$-gon \eqref{eq:GF n to n-k}, we have not yet identified an appropriate transformation that would render it in the form of a rational function.}

Expressing rational numbers with irrational numbers is not unusual in mathematics, with the general term formula of the famous Fibonacci sequence being an example
\begin{equation}
\begin{aligned}
F_n=\frac{1}{\sqrt{5}}\left(\left(\frac{1+\sqrt{5}}{2}\right)^n-\left(\frac{1-\sqrt{5}}{2}\right)^n\right).
\end{aligned}
\end{equation}
This may suggest that in the existing work, the study of the analytic structure of the master integrals or reduction coefficients is perhaps not yet deep enough.

In numerical computations on computers, maintaining a function in rational form allows us to preserve the accuracy of the entire numerical calculation through methods such as \textbf{Finite Field Reconstruction}, ensuring complete precision. When irrational terms are present, we are forced to use floating-point numbers, which can lead to increased computational time and a reduction in precision. However, based on the definition of coefficient $\textbf{C}^{(a'_1,...,a'_k)}_{\{b_1,...,b_{k-1}\}}(n)$ \eqref{eq:define CCC} and the formula of \eqref{eq:define XY}\eqref{eq: define abpq}\eqref{eq:c1 and c2}\eqref{eq:define ABPQ}, it can be seen that the only irrational term appearing in the generating function \eqref{eq:GF n to n-k} from an $n$-gon to an $(n-k)$-gon is \eqref{eq:irr term}. Therefore, by extending the rational number field $\mathbb{Q}$ to the field of form $\mathbb{Q}+\sqrt{X}\cdot \mathbb{Q}$, we can similarly employ finite field reconstruction methods to maintain computational accuracy and save computational time. Nonetheless, we maintain our belief that there may exist a rationalized analytical representation for the generating function of one-loop tensor integral. Investigating methods to derive such a rationalized expression is a meaningful avenue for future research.

Secondly, the generating function we discuss in this paper is in any spacetime dimension $D$ where all scalar integrals with single pole are irreducible. When the spacetime dimension $D$ takes a finite value (such as $4-2\epsilon$), some scalar integrals will become reducible scalar integrals, then what would the form of the generating functions be in this case? Additionally, this paper focuses on one-loop tensor integrals that solely encompass single poles. As for one-loop integrals involving higher poles, what would be the corresponding generating functions for reduction coefficients? We have already computed some results, which will be presented in future articles.

Thirdly, the simplicity of the method in this paper lies in the fact that the recursion relation established is an ordinary differential equation rather than a set of partial differential equations. The foundation of this relation is the study of the analytic structure of Feynman Parameterization in projective space. Could this logic be extrapolated to two loops, thus obtaining the general form of the two-loop generating function, or at least, obtain some simpler recursion relations? In \cite{Kosower:2018obg}, Kosower focus on two-loop integrals, where he only deriving generating functions for the reduction coefficients associated with the most intricate master integral. His approach is applied to only a limited set of concrete examples. In contrast, our goal is to offer a comprehensive and systematic treatment encompassing all generating function for general two-loop integrals. For the case of two-loop diagrams, we have undertaken some initial attempts and obtained preliminary results, which will be presented in future work. Nevertheless, it remains a significant challenge.

	\section*{Acknowledgments}
We would like to thank Prof. Dr. Bo Feng, Dr. Jiaqi Chen and Mr. Xiang Li for useful discussion. We are also grateful to Prof. Dr. Johannes Bluemlein for providing valuable feedback on this paper.

	\begin{appendices}
\section{Solving differential equations}
\label{sec:int}

The standard form of the first-order linear ordinary differential equation is:

\begin{equation}
y' + p(x)y = q(x).
\end{equation}
The solution is

\begin{equation}
y = e^{-\int p(x) \, dx} \int q(x)e^{\int p(x) \, dx} \, dx.
\label{eq:SP 01}
\end{equation}
This result can be used as a formula.

In our cases\eqref{eq:relation 02}, the differential equation has the typical form with a 
non-homogeneous term $g(x)$

\begin{equation}
    \left((a-1)+(a/2)A_1x+A_2x^2\right)y(x)+x(1+A_1x+A_2x^2)y'(x)=g(x),
\end{equation}
where $a,A_1,A_2$ are several coefficients which are independent of the variable $x$. Then $p(x)$, $q(x)$ in\eqref{eq:SP 01}:
\begin{equation}
    \begin{aligned}
p(x)=\frac{(a-1)+(a/2)A_1x+A_2x^2}{x(1+A_1x+A_2x^2)},
 \end{aligned}
\end{equation}

\begin{equation}
    \begin{aligned}
q(x)=\frac{g(x)}{x(A_1+A_2x+A_3x^2)}.
 \end{aligned}
\end{equation}
We split the integral of $p(x)$ into
\begin{equation}
    \begin{aligned}
\int p(x) dx&=\int \frac{(a-1)+(a/2) A_1 x+A_2x^2}{x(1+A_1x+A_2x^2)}\ dx\\
&=\int \bigg(\frac{a-1}{x}+(1-\frac{a}{2})\frac{(A_1+2A_2x)}{(1+A_1x+A_2x^2)}\bigg)\ dx\\
=&\int \bigg(\frac{a-1}{x}+(1-\frac{a}{2})\frac{(1+A_1x+A_2x^2)'}{1+A_1x+A_2x^2}\bigg)\ dx\\
=&(a-1)\ln(x)+(1-\frac{a}{2})\ln(1+A_1x+A_2x^2)+Const.
    \end{aligned}
\end{equation}
Then the general solution of the homogeneous part is

\begin{equation}
    y_H(x)=e^{-\int p(x) \, dx}=Const\cdot  x^{1-a}(1+A_1x+A_2x^2)^{\frac{a-2}{2}}.
\end{equation}
By \eqref{eq:SP 01}, the solution of the original equation is

\begin{equation}
\begin{aligned}
y(x)&= e^{-\int p(x) \, dx} \int q(x)e^{\int p(x) \, dx} \, dx\\
&=x^{1-a}(1+A_1x+A_2x^2)^{\frac{a-2}{2}}\left(\int_0^x  g(u)u^{a-2} (1+ A_1u+A_2u^2)^{-\frac{a}{2}}du+Const\right).
\end{aligned}
\end{equation}
In the case of the reduction of an $n$-gon to an $n$-gon, the non-homogeneous part is a constant $g(x)=(a-1)$. By following integral formula:

\begin{equation}
    \begin{aligned}
    &\int(y-x_{1})^{-\frac{a}{2}}(y-x_{2})^{-\frac{a}{2}}dy=\frac{(y-x_1)^{-\frac{a}{2}}(y-x_{2})^{\frac{2-a}{2}}}{1-a}\cdot \ _2F_1\left(\begin{array}{l}
1,a/2  \\
 a
\end{array} \bigg|\frac{x_{2} - x_{1}}{y - x_1}\right)+Const,
    \end{aligned}
\end{equation}
we can solve
\begin{equation}
    \begin{aligned}
&y(x)
= e^{-\int p(x) \, dx} \int q(x)e^{\int p(x) \, dx} \, dx\\
&=x^{1-a}(1+A_1x+A_2x^2)^{\frac{a-2}{2}}\left((a-1)\int_0^x u^{a} (1+ A_1u+A_2u^2)^{-\frac{a}{2}}u^{-2}du+Const\right)\\
&=\frac{1}{x}\left(\frac{1}{x^2}+A_1\frac{1}{x}+A_2\right)^{\frac{a-2}{2}}\left((a-1)\int_\infty^{\frac{1}{x}}-\left(\frac{1}{u^2}+A_1\frac{1}{u}+A_2\right)^{-\frac{a}{2}}d\left(\frac{1}{u}\right)+Const\right)\\
&=(1-a) \frac{1}{x} \left(\frac{1}{x}-x_{+}\right)^{\frac{a-2}{2}}\left(\frac{1}{x}-x_{-}\right)^{\frac{a-2}{2}}\left(\int_\infty^{\frac{1}{x}}\left(y-x_{+}\right)^{-\frac{a}{2}}\left(y-x_{-}\right)^{-\frac{a}{2}}dy +Const\right)\\
&=\frac{1}{x} \left(\frac{1}{x}-x_{+}\right)^{\frac{a-2}{2}}\left(\frac{1}{x}-x_{-}\right)^{\frac{a-2}{2}}  \\
&\times \left(\Bigg\{(y-x_{+})^{-\frac{a}{2}}(y-x_{-})^{\frac{2-a}{2}}\cdot \ _2F_1\left(\begin{array}{l}
1,a/2  \\
 a
\end{array} \bigg|\frac{x_{-} - x_{+}}{y - x_+}\right)\Bigg|_{y=\infty}^{\frac{1}{x}}+Const\right)\\
&=x^{1-a}(1-x_{+}\cdot x)^{\frac{a-2}{2}}(1-x_{-}\cdot x)^{\frac{a-2}{2}}\cdot Const+\frac{1}{1-x_{+}\cdot x}\cdot \ _2F_1\left(\begin{array}{l}
1,a/2  \\
 a
\end{array} \bigg|\frac{(x_{-} - x_{+})x}{1 - x_+\cdot x}\right),
    \end{aligned}
\end{equation}
where $x_{-},x_{+}$ are two roots of the quadratic equation $x^2+A_1x+A_2=0$,\footnote{We can also choose $x_+=\frac{-A_1-\sqrt{A_1^2-4A_2}}{2},\ x_{-}=\frac{-A_1+\sqrt{A_1^2-4A_2}}{2}$. In this way, we  will have another form of solution. Two forms are equivalent due to the properties of hypergeometric functions.}

\begin{equation}
x_+=\frac{-A_1+\sqrt{A_1^2-4A_2}}{2},\ x_{-}=\frac{-A_1-\sqrt{A_1^2-4A_2}}{2}.
\end{equation}

\section{Numerical check}
\label{sec:num check}

	In this section we will verify our generating function numerically by comparing to the result produced by the cpp version of \verb*|FIRE6|\cite{Smirnov:2019qkx}. The tensor integral we reduce is defined as follows:
	\begin{equation}
		\int\frac{d^D l}{i\pi^{D/2}}\frac{(2R\cdot l)^{r}}{D_1D_2\cdots D_n}.
	\end{equation}

	\subsection{Tadpole}
			Let us first begin with a trivial Tadpole case, where we have only one master integral. We choose the numeric value of the kinetic variable as follows:
	\begin{align}
		&M_1^2\to \frac{5}{4},q_1^2\to 0,R^2\to \frac{133}{10},q_1 \cdot R\to 0,D\to \frac{969}{50}. 
	\end{align}
	
	\begin{itemize}
		\item Tadpole to Tadpole $1\to1$
		 	\begin{table}[htbp]\centering\begin{tabular}{|c|c|c|c|}
	\hline
	\text{Rank} & 0 & 1 & 2 \\
	\hline
	\text{Result} & 1. & 0. & 3.4313725490196076 \\
	\hline
	\text{Rank} & 3 & 4 & 5 \\
	\hline
	\text{Result} & 0. & 32.0186540472129 & 0. \\
	\hline
	\text{Rank} & 6 & 7 & 8 \\
	\hline
	\text{Result} & 455.355109952878 & 0. & 8351.765314541557 \\
	\hline
	\text{Rank} & 9 & 10 & 11 \\
	\hline
	\text{Result} & 0. & 182561.41492889417 & 0. \\
	\hline
			\end{tabular}\caption{Reduction Result of Tadpole to Tadpole by \texttt{FIRE6}}\end{table}
		
	The numerical Taylor expansion of the generating function is given as follows: 
		\begin{align}
			\textbf{GF}{}_{1\to1}(\mathcal{t})&=1+ 3.4313725490196076\cdot\mathcal{t}^2 + 32.0186540472129\cdot \mathcal{t}^4 \notag\\&
			+ 455.355109952878\cdot \mathcal{t}^6 + 8351.765314541557\cdot\mathcal{t}^8\notag\\&+ 182561.41492889417\cdot\mathcal{t}^{10}+\mathcal{O}(\mathcal{t}^{12}),
		\end{align}
		which match the result of \verb*|FIRE6| perfectly. 
	\end{itemize}

	\subsection{Bubble}
			We choose the numeric value of the kinetic variable as follows:
	\begin{align}
		&M_1^2\to \frac{5}{4},M_2^2\to \frac{64}{25},q_1^2\to 0,q_2^2\to \frac{69}{50},q_1\cdot q_2\to 0,\notag\\
		&R^2\to \frac{133}{10},q_1\cdot R\to 0,q_2\cdot R\to \frac{367}{100},D\to \frac{969}{50}. 
	\end{align}
	For a tensor Bubble integral, we have totally 3 master integrals: 
	\begin{itemize}
		\item Bubble to Bubble $2\to2$
		 
		 	\begin{table}[htbp]\centering\begin{tabular}{|c|c|c|c|}
		 			\hline
		 			\text{Rank} & 0 & 1 & 2 \\
		 			\hline
		 			\text{Result} & 1. & 0.18615942028985508 & 0.9969550236277754 \\
		 			\hline
		 			\text{Rank} & 3 & 4 & 5 \\
		 			\hline
		 			\text{Result} & 0.5438748755636772 & 2.706729663300785 & 2.394357906464815 \\
		 			\hline
		 			\text{Rank} & 6 & 7 & 8 \\
		 			\hline
		 			\text{Result} & 11.220093828863014 & 13.471511680729403 & 60.095504925096684 \\
		 			\hline
		 			\text{Rank} & 9 & 10 & 11 \\
		 			\hline
		 			\text{Result} & 89.67183068422794 & 384.30945689202724 & 675.8039722773336 \\
		 			\hline
		 	\end{tabular}\caption{Reduction Result of Bubble to Bubble by \texttt{FIRE6}}\end{table}
		 The numerical Taylor expansion of the generating function is given as follows: 
		 	\begin{align}
		 	\textbf{GF}{}_{2\to2}&(\mathcal{t})=1 + 0.18615942028985508\cdot\mathcal{t} + 0.9969550236277754\cdot\mathcal{t}^2 \notag\\
		 	&+ 0.5438748755637103\cdot\mathcal{t}^3 + 2.7067296633008473\cdot\mathcal{t}^4 \notag\\
		 	&+ 2.394357906466195\cdot\mathcal{t}^5 + 11.220093828873562\cdot\mathcal{t}^6 \notag\\
		 	&+ 13.471511680934789\cdot\mathcal{t}^7 + 60.09550492544111\cdot\mathcal{t}^8 \notag\\
		 	&+ 89.67183069722255\cdot\mathcal{t}^9 + 384.3094569884708\cdot\mathcal{t}^{10} \notag\\
		 	&+ 675.8039743685777\cdot\mathcal{t}^{11}+\mathcal{O}(\mathcal{t}^{12}).
		 \end{align}
		 
		\item Bubble to Tadpole $2\to1$
		
		There are two Tadpole master integrals in this case. 
			\begin{table}[htbp]\centering\begin{tabular}{|c|c|c|c|}
						\hline
						\text{Rank} & 0 & 1 & 2 \\
						\hline
						\text{Result} & 0. & 2.6594202898550723 & 20.390645089351526 \\
						\hline
						\text{Rank} & 3 & 4 & 5 \\
						\hline
						\text{Result} & 175.8509958812531 & 1611.1537585017563 & 15713.736765104697 \\
						\hline
						\text{Rank} & 6 & 7 & 8 \\
						\hline
						\text{Result} & 161200.4597648606 & \text{$1.726292310058\times 10^6$} &
						\text{$1.91794043011\times 10^7$} \\
						\hline
						\text{Rank} & 9 & 10 & 11 \\
						\hline
						\text{Result} & \text{$2.20010455418\times 10^8$} & \text{$2.59573008468\times 10^9$} &
						\text{$3.13986818427\times 10^{10}$} \\
						\hline
			\end{tabular}\caption{Reduction Result of Bubble to $D_2$ Tadpole by \texttt{FIRE6}}\end{table}
	
		The numerical Taylor expansion of the generating function is given as follows,
			 	\begin{align}
				\textbf{GF}{}_{2\to1;\widehat{1}}&(\mathcal{t})=
			 	   2.6594202898\cdot\mathcal{t} + 20.390645089351\cdot\mathcal{t}^2 +175.85099588125\cdot\mathcal{t}^3 \notag\\
			   &+1611.153758502\cdot\mathcal{t}^4 +15713.736765105\cdot\mathcal{t}^5 +161200.45976486\cdot\mathcal{t}^6 \notag\\
			    &  +1.72629231\times 10^6\cdot\mathcal{t}^7 +1.91794043\times 10^7\cdot\mathcal{t}^8 +2.20010455\times 10^8\cdot\mathcal{t}^9 \notag\\
			     & +2.595730085\times 10^9\cdot\mathcal{t}^{10} +3.13986818\times 10^{10}\cdot\mathcal{t}^{11}+\mathcal{O}(\mathcal{t}^{12}).
			\end{align}
		
		\begin{table}[htbp]\centering\begin{tabular}{|c|c|c|c|}
					\hline
					\text{Rank} & 0 & 1 & 2 \\
					\hline
					\text{Result} & 0. & -2.6594202898550723 & -0.4853067411154148 \\
					\hline
					\text{Rank} & 3 & 4 & 5 \\
					\hline
					\text{Result} & -14.069834047182031 & -3.9250909295456946 & -132.48553286333672 \\
					\hline
					\text{Rank} & 6 & 7 & 8 \\
					\hline
					\text{Result} & -40.82736337173687 & -1820.708411154273 & -554.9841273406652 \\
					\hline
					\text{Rank} & 9 & 10 & 11 \\
					\hline
					\text{Result} & -32477.51478825142 & -9528.7663298311 & -697318.0606970708 \\
					\hline
		\end{tabular}\caption{Reduction Result of Bubble to $D_1$ Tadpole by \texttt{FIRE6}}\end{table}
	
		The numerical Taylor expansion of the generating function is given as follows: 
	 	\begin{align}
		\textbf{GF}{}_{2\to1;\widehat{2}}&(\mathcal{t})=
		        -2.6594202898550\cdot\mathcal{t} - 0.4853067411154\cdot\mathcal{t}^2 - 14.0698340471820\cdot\mathcal{t}^3 \notag\\
		   &    - 3.9250909295456\cdot\mathcal{t}^4 - 132.4855328633367\cdot\mathcal{t}^5 - 40.8273633717368\cdot\mathcal{t}^6 \notag\\
		   & - 1820.7084111542733\cdot\mathcal{t}^7 - 554.9841273406652\cdot\mathcal{t}^8 - 32477.5147882514220\cdot\mathcal{t}^9 \notag\\
	       &	- 9528.7663298311876\cdot\mathcal{t}^{10} - 697318.0606970731579\cdot\mathcal{t}^{11}+\mathcal{O}(\mathcal{t}^{12}).
	\end{align}
		
	\end{itemize}
	
	\subsection{Triangle}
		We choose the numeric value of the kinetic variable as follows:
	\begin{align}
		&M_1^2\to \frac{5}{4},M_2^2\to \frac{64}{25},M_3^2\to \frac{357}{100},q_1^2\to 0,q_2^2\to \frac{69}{50},q_3^2\to\frac{129}{25},q_1 \cdot q_2\to 0,q_1\cdot q_3\to 0,\notag\\
		&q_2\cdot q_3\to \frac{41}{20},R^2\to \frac{133}{10},q_1 \cdot R\to 0,q_2\cdot R\to \frac{367}{100},q_3 \cdot R\to \frac{89}{10},D\to \frac{969}{50}. 
	\end{align}
	
	\begin{itemize}
		\item Triangle to Triangle,$3\to3$
		\begin{table}[htbp]\centering\begin{tabular}{|c|c|c|c|}
			\hline
			\text{Rank} & 0 & 1 & 2 \\
			\hline
			\text{Result} & 1. & 4.647429667957373 & 21.424063591070208 \\
			\hline
			\text{Rank} & 3 & 4 & 5 \\
			\hline
			\text{Result} & 97.94451396121778 & 443.9628092202622 & 1994.7300798765605 \\
			\hline
			\text{Rank} & 6 & 7 & 8 \\
			\hline
			\text{Result} & 8880.900093335815 & 39165.84906934403 & 171022.14771781594 \\
			\hline
			\text{Rank} & 9 & 10 & 11 \\
			\hline
			\text{Result} & 739055.1488881096 & \text{$3.15884655856376\times 10^6$} &
			\text{$1.3344673752179299\times 10^7$} \\
			\hline
		\end{tabular}\caption{Reduction Result of Triangle to Triangle by \texttt{FIRE6}}\end{table}
	
		The numerical Taylor expansion of the generating function is given as follows: 
		\begin{align}
		\textbf{GF}&{}_{3\to3}(\mathcal{t})=
		1. + 4.647429667957\cdot\mathcal{t} + 21.42406359107\cdot\mathcal{t}^2 + 97.9445139612\cdot\mathcal{t}^3 \notag\\
		&+ 443.9628092202622\cdot\mathcal{t}^4 + 1994.7300798765605\cdot\mathcal{t}^5 + 8880.900093335815\cdot\mathcal{t}^6 \notag\\
		&+ 39165.84906934403\cdot\mathcal{t}^7 + 171022.14771781594\cdot\mathcal{t}^8 + 739055.1488881096\cdot\mathcal{t}^9 \notag\\
	&	+ 3.15884655856376\times 10^6\cdot\mathcal{t}^{10} + 1.3344673752179299\times 10^7\cdot\mathcal{t}^{11}+\mathcal{O}(\mathcal{t}^{12}).
	\end{align}
	
		\item Triangle to Bubble,$3\to2$
		
		There are 3 Bubble master integrals in this case, we choose two typical integrals to verify our generating function. 
			\begin{table}[htbp]\centering\begin{tabular}{|c|c|c|c|}
				\hline
				\text{Rank} & 0 & 1 & 2 \\
				\hline
				\text{Result} & 0. & 1.867765479902683 & 27.80579292856186 \\
				\hline
				\text{Rank} & 3 & 4 & 5 \\
				\hline
				\text{Result} & 326.58851259528217 & 3575.377043547329 & 38296.692771726164 \\
				\hline
				\text{Rank} & 6 & 7 & 8 \\
				\hline
				\text{Result} & 408802.874852242 & \text{$4.3817950217888\times 10^6$} &
				\text{$4.730368560885\times 10^7$} \\
				\hline
				\text{Rank} & 9 & 10 & 11 \\
				\hline
				\text{Result} & \text{$5.148927898936\times 10^8$} & \text{$5.6522910727\times 10^9$} &
				\text{$6.2570409303155\times 10^{10}$} \\
				\hline
			\end{tabular}\caption{Reduction Result of Triangle to $D_2D_3$ Bubble by \texttt{FIRE6}}\end{table}
	
		The numerical Taylor expansion of the generating function is given as follows: 
		\begin{align}
			&\textbf{GF}{}_{3\to2;\widehat{1}}(\mathcal{t})=
			      1.86776547990\cdot\mathcal{t} +27.80579292856\cdot\mathcal{t}^2 +326.58851259528\cdot\mathcal{t}^3 \notag\\
			 &+ 3575.377043547329\cdot\mathcal{t}^4 +38296.692771726167\cdot\mathcal{t}^5 + 408802.87485224194\cdot\mathcal{t}^6 \notag\\
			  & + 4.3817950217888\times 10^6\cdot \mathcal{t}^7 + 4.730368560885\times 10^7\cdot \mathcal{t}^8  
			+ 5.1489278989\times 10^8\cdot \mathcal{t}^9\notag\\&+ 5.6522910727\times 10^9\cdot \mathcal{t}^{10} + 6.2570409303\times 10^{10}\cdot\mathcal{t}^{11}+\mathcal{O}(\mathcal{t}^{12}).
		\end{align}
		
			\begin{table}[htbp]\centering\begin{tabular}{|c|c|c|c|}
			\hline
			\text{Rank} & 0 & 1 & 2 \\
			\hline
			\text{Result} & 0. & -0.2371928862694034 & -2.039991135700739 \\
			\hline
			\text{Rank} & 3 & 4 & 5 \\
			\hline
			\text{Result} & -12.812290548944913 & -69.25281622582993 & -336.83725868345635 \\
			\hline
			\text{Rank} & 6 & 7 & 8 \\
			\hline
			\text{Result} & -1487.8795033941478 & -5880.3689489331 & -19650.58517698222 \\
			\hline
			\text{Rank} & 9 & 10 & 11 \\
			\hline
			\text{Result} & -44502.51591828201 & 48219.21614887317 & \text{$1.5017448726269251\times 10^6$} \\
			\hline
		\end{tabular}\caption{Reduction Result of Triangle to $D_1D_3$ Bubble by \texttt{FIRE6}}\end{table}
	
		The numerical Taylor expansion of the generating function is given as follows: 
	\begin{align}
		\textbf{GF}&{}_{3\to2;\widehat{2}}(\mathcal{t})=
		      -0.2371928863\cdot \mathcal{t} - 2.0399911357\cdot\mathcal{t}^2 - 12.81229054894\cdot\mathcal{t}^3 \notag\\
		    &- 69.25281622583\cdot\mathcal{t}^4 - 336.83725868346\cdot\mathcal{t}^5 - 1487.87950339415\cdot\mathcal{t}^6 \notag\\
		  &- 5880.368948933\cdot\mathcal{t}^7 -  19650.5851769822\cdot\mathcal{t}^8 -  44502.51591828201\cdot\mathcal{t}^9 \notag\\
		&+  48219.2161488732\cdot\mathcal{t}^{10} + 1.50174487263\times 10^6\cdot\mathcal{t}^{11}+\mathcal{O}(\mathcal{t}^{12}).
	\end{align}
	
		\item Triangle to Tadpole, $3\to1$
		
		There are 3 Tadpole master integrals in this case.
			\begin{table}[htbp]\centering\begin{tabular}{|c|c|c|c|}
				\hline
				\text{Rank} & 0 & 1 & 2 \\
				\hline
				\text{Result} & 0. & 0. & 3.5943363780985007 \\
				\hline
				\text{Rank} & 3 & 4 & 5 \\
				\hline
				\text{Result} & 120.15405371902189 & 2856.4892399384603 & 60343.814251369666 \\
				\hline
				\text{Rank} & 6 & 7 & 8 \\
				\hline
				\text{Result} & \text{$1.22027431911\times 10^6$} & \text{$2.43997498612\times 10^7$} &
				\text{$4.899382591712\times 10^8$} \\
				\hline
				\text{Rank} & 9 & 10 & 11 \\
				\hline
				\text{Result} & \text{$9.95303381377\times 10^9$} & \text{$2.052398465613\times 10^{11}$} &
				\text{$4.300770011403\times10^{12}$} \\
				\hline
			\end{tabular}\caption{Reduction Result of Triangle to $D_3$ Tadpole by \texttt{FIRE6}}\end{table}
	
		The numerical Taylor expansion of the generating function is given as follows: 
			\begin{align}
				\textbf{GF}&{}_{3\to1;\widehat{12}}(\mathcal{t})=
				      3.594336378098\cdot\mathcal{t}^2 + 120.154053719\cdot\mathcal{t}^3 + 2856.489239938\cdot\mathcal{t}^4 \notag\\
				&+ 60343.81425137\cdot\mathcal{t}^5 + 1.220274319\times 10^6\cdot\mathcal{t}^6 + 2.439974986\times 10^7\cdot\mathcal{t}^7 \notag\\
				 &   + 4.899382592\times 10^8\cdot\mathcal{t}^8 + 9.953033814\times 10^9\cdot\mathcal{t}^9  + 2.052398466\times 10^{11}\cdot\mathcal{t}^{10 } \notag\\
				  & +4.30077001\times 10^{12}\cdot\mathcal{t}^{11}+\mathcal{O}(\mathcal{t}^{12}).
			\end{align}
			
			\begin{table}[htbp]\centering\begin{tabular}{|c|c|c|c|}
				\hline
				\text{Rank} & 0 & 1 & 2 \\
				\hline
				\text{Result} & 0. & 0. & 4.745489600791057 \\
				\hline
				\text{Rank} & 3 & 4 & 5 \\
				\hline
				\text{Result} & 24.912564026958766 & 142.5977420103585 & 671.4679047318388 \\
				\hline
				\text{Rank} & 6 & 7 & 8 \\
				\hline
				\text{Result} & 3235.5137500307833 & 14241.416319020253 & 63070.04241030086 \\
				\hline
				\text{Rank} & 9 & 10 & 11 \\
				\hline
				\text{Result} & 258765.71357130772 & \text{$1.06056655995\times 10^6$} &
				\text{$3.96215910911\times 10^6$} \\
				\hline
			\end{tabular}\caption{Reduction Result of Triangle to $D_1$ Tadpole by \texttt{FIRE6}}\end{table}
	
		The numerical Taylor expansion of the generating function is given as follows: 
			\begin{align}
				\textbf{GF}&{}_{3\to1;\widehat{23}}(\mathcal{t})=
				      4.74548960079\cdot\mathcal{t}^2 + 24.91256402696\cdot\mathcal{t}^3 + 142.59774201036\cdot\mathcal{t}^4 \notag\\
				  & + 671.4679047318\cdot\mathcal{t}^5 + 3235.5137500308\cdot\mathcal{t}^6 + 14241.4163190203\cdot\mathcal{t}^7 \notag\\
				 &+ 63070.0424103009\cdot\mathcal{t}^8 + 258765.71357130772\cdot\mathcal{t}^9 + 1.0605665599514\times 10^6\cdot \mathcal{t}^{10} \notag\\
				    & + 3.9621591091\times 10^6\cdot\mathcal{t}^{11}+\mathcal{O}(\mathcal{t}^{12}).
			\end{align}
		
			\begin{table}[htbp]\centering\begin{tabular}{|c|c|c|c|}
			\hline
			\text{Rank} & 0 & 1 & 2 \\
			\hline
			\text{Result} & 0. & 0. & -8.339825978889557 \\
			\hline
			\text{Rank} & 3 & 4 & 5 \\
			\hline
			\text{Result} & -142.45218080322186 & -1909.4499907718186 & -23603.993924573977 \\
			\hline
			\text{Rank} & 6 & 7 & 8 \\
			\hline
			\text{Result} & -282767.93446188176 & \text{$-3.347293948348\times 10^6$} &
			\text{$-3.9512933931\times 10^7$} \\
			\hline
			\text{Rank} & 9 & 10 & 11 \\
			\hline
			\text{Result} & \text{$-4.673481474464\times 10^8$} & \text{$-5.55382934594\times 10^9$} &
			\text{$-6.64265408461\times 10^{10}$} \\
			\hline
		\end{tabular}\caption{Reduction Result of Triangle to $D_2$ Tadpole by \texttt{FIRE6}}\end{table}
	
		The numerical Taylor expansion of the generating function is given as follows: 
			\begin{align}
				\textbf{GF}&{}_{3\to1;\widehat{13}}(\mathcal{t})=
				      -8.33982597889\cdot\mathcal{t}^2 - 142.45218080322\cdot\mathcal{t}^3 - 1909.44999077182\cdot\mathcal{t}^4 \notag\\
				 &- 23603.993924574\cdot\mathcal{t}^5 - 282767.9344618818\cdot\mathcal{t}^6 - 3.34729394835\times 10^6\cdot\mathcal{t}^7\notag\\ 
				  &   - 3.95129339314\times 10^7\cdot\mathcal{t}^8 - 4.67348147446\times 10^8\cdot\mathcal{t}^9  \notag\\
				   & - 5.55382934594\times 10^9\cdot\mathcal{t}^{10} - 6.64265408461\times 10^{10}\cdot\mathcal{t}^{11}+\mathcal{O}(\mathcal{t}^{12}).
			\end{align}
\end{itemize}

	\subsection{Box}
	We choose the numeric value of the kinetic variable as follows:
\begin{align}
	&M_1^2\to \frac{5}{4},M_2^2\to \frac{64}{25},M_3^2\to \frac{357}{100},M_4^2\to
	\frac{339}{50},q_1^2\to 0,q_2^2\to \frac{69}{50},q_3^2\to
	\frac{129}{25},\notag\\
	&q_4^2\to \frac{2069}{100},q_1\cdot q_2\to 0,q_1\cdot
	q_3\to 0,q_1\cdot q_4\to 0,q_2\cdot q_3\to \frac{41}{20},\notag\\
	&q_2\cdot q_4\to
	\frac{93}{25},q_3\cdot q_4\to \frac{56}{5},R^2\to \frac{133}{10},q_1\cdot R\to 0,\notag\\
	&q_2\cdot
	R\to \frac{367}{100},q_3\cdot R\to \frac{89}{10},q_4\cdot R\to \frac{2413}{100},D\to \frac{969}{50}. 
\end{align}
		
		\begin{itemize}
			
		\item Box to Box,$4\to4$
	
		\begin{table}[htbp]\centering\begin{tabular}{|c|c|c|c|}
			\hline
			\text{Rank} & 0 & 1 & 2 \\
			\hline
			\text{Result} & 1. & -3.6920841521541408 & 16.55377492166991 \\
			\hline
			\text{Rank} & 3 & 4 & 5 \\
			\hline
			\text{Result} & -82.69660780758952 & 447.6598650239185 & -2578.2813404112676 \\
			\hline
			\text{Rank} & 6 & 7 & 8 \\
			\hline
			\text{Result} & 15614.703573346487 & -98601.17919296624 & 645114.4523761669 \\
			\hline
			\text{Rank} & 9 & 10 & 11 \\
			\hline
			\text{Result} & \text{$-4.35187298694\times 10^6$} & \text{$3.0151991318039\times 10^7$} &
			\text{$-2.138897834607\times 10^8$} \\
			\hline
		\end{tabular}\caption{Reduction Result of Box to Box by \texttt{FIRE6}}\end{table}
	
	The numerical Taylor expansion of the generating function is given as follows: 
	    \begin{align}
		\textbf{GF}&{}_{4\to4}(\mathcal{t})=
		1. - 3.692084152154\cdot\mathcal{t} + 16.5537749217\cdot\mathcal{t}^2 - 82.6966078076\cdot\mathcal{t}^3 \notag\\
		&+ 447.6598650239\cdot\mathcal{t}^4 -2578.281340411\cdot\mathcal{t}^5 + 15614.70357335\cdot\mathcal{t}^6 \notag\\
		&-98601.179192966\cdot\mathcal{t}^7 + 645114.4523762\cdot\mathcal{t}^8 - 4.351872986943\times 10^6\cdot\mathcal{t}^9 \notag\\
		&+ 3.015199131804\times 10^7\cdot\mathcal{t}^{10} - 2.1388978346\times 10^8\cdot\mathcal{t}^{11}+\mathcal{O}(\mathcal{t}^{12}).
	\end{align}
	
		\item Box to Triangle, $4\to3$
		
		We choose two typical master integrals to verify our generating function. 
			\begin{table}[htbp]\centering\begin{tabular}{|c|c|c|c|}
				\hline
				\text{Rank} & 0 & 1 & 2 \\
				\hline
				\text{Result} & 0. & 2.297291695479869 & -3.938379081474582 \\
				\hline
				\text{Rank} & 3 & 4 & 5 \\
				\hline
				\text{Result} & 60.596241405397464 & -100.49970077717936 & 2298.7596822646256 \\
				\hline
				\text{Rank} & 6 & 7 & 8 \\
				\hline
				\text{Result} & -1488.7266429821102 & 114099.44338526356 & 128454.40427289286 \\
				\hline
				\text{Rank} & 9 & 10 & 11 \\
				\hline
				\text{Result} & \text{$7.110492574192\times 10^6$} & \text{$2.395974442872\times 10^7$} &
				\text{$5.39796222751\times 10^8$} \\
				\hline
			\end{tabular}\caption{Reduction Result of Box to $D_2D_3D_4$ Triangle by \texttt{FIRE6}}\end{table}
	
		The numerical Taylor expansion of the generating function is given as follows: 
		\begin{align}
			&\textbf{GF}{}_{4\to3;\widehat{1}}(\mathcal{t})=
			   2.29729169548\cdot\mathcal{t} - 3.93837908147\cdot\mathcal{t}^2 + 60.5962414054\cdot\mathcal{t}^3 \notag\\
			&- 100.499700777\cdot\mathcal{t}^4 + 2298.7596822646\cdot\mathcal{t}^5 - 1488.7266429821\cdot\mathcal{t}^6 \notag\\
		  &+114099.44338526357\cdot\mathcal{t}^7 + 128454.4042728928\cdot\mathcal{t}^8 + 7.11049257419\times 10^6\cdot\mathcal{t}^9 \notag\\
		   &   + 2.39597444287\times 10^7\cdot\mathcal{t}^{10 }+ 5.3979622275\times 10^8\cdot\mathcal{t}^{11}+\mathcal{O}(\mathcal{t}^{12}).
		\end{align}
		
			\begin{table}[htbp]\centering\begin{tabular}{|c|c|c|c|}
			\hline
			\text{Rank} & 0 & 1 & 2 \\
			\hline
			\text{Result} & 0. & 1.1727646483063998 & -13.437042633643701 \\
			\hline
			\text{Rank} & 3 & 4 & 5 \\
			\hline
			\text{Result} & 133.44969733467502 & -1305.8124781550728 & 12995.470486241737 \\
			\hline
			\text{Rank} & 6 & 7 & 8 \\
			\hline
			\text{Result} & -132770.84174232915 & \text{$1.395549705737\times 10^6$} &
			\text{$-1.508260416218\times 10^7$} \\
			\hline
			\text{Rank} & 9 & 10 & 11 \\
			\hline
			\text{Result} & \text{$1.673393577774\times 10^8$} & \text{$-1.902215247513\times 10^9$} &
			\text{$2.2110105646826\times 10^{10}$} \\
			\hline
		\end{tabular}\caption{Reduction Result of Box to $D_1D_3D_4$ Triangle by \texttt{FIRE6}}\end{table}
	
		The numerical Taylor expansion of the generating function is given as follows: 
		\begin{align}
		\textbf{GF}&{}_{4\to3;\widehat{2}}(\mathcal{t})=
		    1.172764648\cdot\mathcal{t} - 13.437042634\cdot\mathcal{t}^2 +133.449697335\cdot\mathcal{t}^3 \notag\\
		&-1305.812478155\cdot\mathcal{t}^4 +12995.470486242\cdot\mathcal{t}^5 -132770.84174233\cdot\mathcal{t}^6 \notag\\
	      & +1.3955497057\times 10^6\cdot\mathcal{t}^7 -1.508260416\times 10^7\cdot\mathcal{t}^8 +1.673393578\times 10^8\cdot\mathcal{t}^9 \notag\\
	       &-1.9022152475\times 10^9\cdot\mathcal{t}^{10 }+2.2110105647\times 10^{10}\cdot\mathcal{t}^{11}+\mathcal{O}(\mathcal{t}^{12}).
	\end{align}
	
		\item Box to Bubble, $4\to2$
		
		We choose two typical integrals to verify our generating function. 
			\begin{table}[htbp]\centering\begin{tabular}{|c|c|c|c|}
				\hline
				\text{Rank} & 0 & 1 & 2 \\
				\hline
				\text{Result} & 0. & 0. & 14.283644489205553 \\
				\hline
				\text{Rank} & 3 & 4 & 5 \\
				\hline
				\text{Result} & 204.56675272706224 & 4109.787640764211 & 54424.224763862854 \\
				\hline
				\text{Rank} & 6 & 7 & 8 \\
				\hline
				\text{Result} & 614054.5523170701 & \text{$1.9047172844007\times 10^6$} &
				\text{$-8.985743083129\times 10^7$} \\
				\hline
				\text{Rank} & 9 & 10 & 11 \\
				\hline
				\text{Result} & \text{$-2.6301936458306\times 10^9$} & \text{$-3.119325555871\times 10^{10}$}
				& \text{$1.6840989924241\times 10^{11}$} \\
				\hline
			\end{tabular}\caption{Reduction Result of Box to $D_3D_4$ Bubble by \texttt{FIRE6}}\end{table}
	
		The numerical Taylor expansion of the generating function is given as follows: 
		\begin{align}
			\textbf{GF}&{}_{4\to2;\widehat{12}}(\mathcal{t})=
			  14.283644489\cdot\mathcal{t}^2+204.566752727\cdot\mathcal{t}^3 + 4109.7876407642\cdot\mathcal{t}^4 \notag\\
			  &+ 54424.2247638628\cdot\mathcal{t}^5 + 614054.55231707\cdot\mathcal{t}^6 +   1.904717284\times 10^6\cdot\mathcal{t}^7 \notag\\
			  &- 8.985743083\times 10^7\cdot\mathcal{t}^8 - 2.630193646\times 10^9\cdot\mathcal{t}^9 -3.1193256\times 10^{10}\cdot\mathcal{t}^{10 }\notag\\
			  &+1.6840989924\times 10^{11}\cdot\mathcal{t}^{11}+\mathcal{O}(\mathcal{t}^{12}).
		\end{align}
		
			\begin{table}[htbp]\centering\begin{tabular}{|c|c|c|c|}
			\hline
			\text{Rank} & 0 & 1 & 2 \\
			\hline
			\text{Result} & 0. & 0. & -9.695886285967024 \\
			\hline
			\text{Rank} & 3 & 4 & 5 \\
			\hline
			\text{Result} & -188.13277462466073 & -6091.362674944455 & -152379.17835453493 \\
			\hline
			\text{Rank} & 6 & 7 & 8 \\
			\hline
			\text{Result} & \text{$-4.04931789159\times 10^6$} & \text{$-1.034624729484\times 10^8$} &
			\text{$-2.669747661571\times 10^9$} \\
			\hline
			\text{Rank} & 9 & 10 & 11 \\
			\hline
			\text{Result} & \text{$-6.863487054848\times 10^{10}$} & \text{$-1.7752858181645\times 10^{12}$}
			& \text{$-4.6131078626\times 10^{13}$} \\
			\hline
		\end{tabular}\caption{Reduction Result of Box to $D_1D_4$ Bubble by \texttt{FIRE6}}\end{table}
	
		The numerical Taylor expansion of the generating function is given as follows: 
		\begin{align}
		\textbf{GF}&{}_{4\to2;\widehat{23}}(\mathcal{t})=
		      -9.6958862859\cdot\mathcal{t}^2 -188.1327746246\cdot\mathcal{t}^3 -6091.3626749444\cdot\mathcal{t}^4 \notag\\
		&- 152379.1783545\cdot\mathcal{t}^5 - 4.04931789\times 10^6\cdot\mathcal{t}^6 -1.034624729484\times 10^8\cdot\mathcal{t}^7 \notag\\
		 &    - 2.66974766\times 10^9\cdot\mathcal{t}^8 - 6.863487055\times 10^{10}\cdot\mathcal{t}^9 -1.77528582\times 10^{12}\cdot\mathcal{t}^{10} \notag\\ &- 4.61310786\times 10^{13}\cdot\mathcal{t}^{11}+\mathcal{O}(\mathcal{t}^{12}).
	\end{align}
	
		\item Box to Tadpole $4\to1$
		
		We choose two typical integrals to verify our generating function. 
			\begin{table}[htbp]\centering\begin{tabular}{|c|c|c|c|}
						\hline
						\text{Rank} & 0 & 1 & 2 \\
						\hline
						\text{Result} & 0. & 0. & 0. \\
						\hline
						\text{Rank} & 3 & 4 & 5 \\
						\hline
						\text{Result} & -2.39789456182502 & -26.096086956520185 & 986.8535842106996 \\
						\hline
						\text{Rank} & 6 & 7 & 8 \\
						\hline
						\text{Result} & 53824.37061112356 & \text{$1.945611485647\times 10^6$} &
						\text{$5.97872946799\times 10^7$} \\
						\hline
						\text{Rank} & 9 & 10 & 11 \\
						\hline
						\text{Result} & \text{$1.726812691986\times 10^9$} & \text{$4.81514011073\times 10^{10}$} &
						\text{$1.320524816629\times 10^{12}$} \\
						\hline
			\end{tabular}\caption{Reduction Result of Box to $D_1$ Tadpole by \texttt{FIRE6}}\end{table}
	
		The numerical Taylor expansion of the generating function is given as follows: 
		\begin{align}
			\textbf{GF}{}&_{4\to1;\widehat{234}}(\mathcal{t})=
		        -2.397894561\cdot\mathcal{t}^3 -26.096086956\cdot\mathcal{t}^4 +986.853584210\cdot\mathcal{t}^5 \notag\\
			&+53824.370611123\cdot\mathcal{t}^6 + 1.94561148\times 10^6\cdot\mathcal{t}^7 + 5.97872947\times 10^7\cdot\mathcal{t}^8 \notag\\
			 &  + 1.726813\times 10^9\cdot\mathcal{t}^9 +    4.81514\times 10^{10}\cdot\mathcal{t}^{10 }+ 1.320525\times 10^{12}\cdot\mathcal{t}^{11}+\mathcal{O}(\mathcal{t}^{12}).
		\end{align}
		
			\begin{table}[htbp]\centering\begin{tabular}{|c|c|c|c|}
	\hline
	\text{Rank} & 0 & 1 & 2 \\
	\hline
	\text{Result} & 0. & 0. & 0. \\
	\hline
	\text{Rank} & 3 & 4 & 5 \\
	\hline
	\text{Result} & 21.912939954259137 & 780.2207801551945 & 20884.548959274303 \\
	\hline
	\text{Rank} & 6 & 7 & 8 \\
	\hline
	\text{Result} & 490764.9683116434 & \text{$1.078965830487\times 10^7$} &
	\text{$2.265411104988\times 10^8$} \\
	\hline
	\text{Rank} & 9 & 10 & 11 \\
	\hline
	\text{Result} & \text{$4.57159700668\times 10^9$} & \text{$8.82631887425\times 10^{10}$} &
	\text{$1.6031881275156\times 10^{12}$} \\
	\hline
		\end{tabular}\caption{Reduction Result of Box to $D_2$ Tadpole by \texttt{FIRE6}}\end{table}
	
		The numerical Taylor expansion of the generating function is given as follows: 
		\begin{align}
			\textbf{GF}&{}_{4\to1;\widehat{134}}(\mathcal{t})=
 21.912939954\cdot\mathcal{t}^3 +780.220780155\cdot\mathcal{t}^4 +20884.54895927\cdot\mathcal{t}^5 \notag\\
&+490764.9683116\cdot\mathcal{t}^6 +1.0789658\times 10^7\cdot\mathcal{t}^7 +2.2654111\times 10^8\cdot\mathcal{t}^8 \notag\\
&+4.571597\times 10^9\cdot\mathcal{t}^9 +8.826319\times 10^{10}\cdot\mathcal{t}^{10 }+1.6031881\times 10^{13}\cdot\mathcal{t}^{11}+\mathcal{O}(\mathcal{t}^{12}).
		\end{align}
	\end{itemize}

	\subsection{Pentagon}
		We choose the numeric value of the kinetic variable as follows:
	\begin{align}
		&M_1^2\to \frac{5}{4},M_2^2\to \frac{64}{25},M_3^2\to \frac{357}{100},M_4^2\to
		\frac{339}{50},M_5^2\to \frac{489}{100},q_1^2\to 0,q_2^2\to \frac{69}{50},q_3^2\to
		\frac{129}{25},\notag\\
		&q_4^2\to \frac{2069}{100},q_5^2\to \frac{1957}{25},q_1\cdot q_2\to 0,q_1\cdot
		q_3\to 0,q_1\cdot q_4\to 0,q_1\cdot q_5\to 0,q_2\cdot q_3\to \frac{41}{20},\notag\\
		&q_2\cdot q_4\to
		\frac{93}{25},q_2\cdot q_5\to \frac{721}{100},q_3\cdot q_4\to \frac{56}{5},q_3\cdot q_5\to
		\frac{473}{20},q_4\cdot q_5\to \frac{4437}{100},R^2\to \frac{133}{10},\notag\\
		&q_1\cdot R\to 0, q_2\cdot
		R\to \frac{367}{100},q_3\cdot R\to \frac{89}{10},q_4\cdot R\to \frac{2413}{100},q_5\cdot R\to
		\frac{3737}{100},D\to \frac{969}{50}. 
	\end{align}
	\begin{itemize}
		\item Pentagon to Pentagon $5\to 5$
	
			\begin{table}[htbp]\centering
				\begin{tabular}{|c|c|c|c|}
					\hline
					\text{Rank} & 0 & 1 & 2 \\
					\hline
					\text{Result} & 1. & 28.14374534067907 & 891.8998182261959 \\
					\hline
					\text{Rank} & 3 & 4 & 5 \\
					\hline
					\text{Result} & 30720.54870031494 & \text{1.1282643383555224*${}^{\wedge}$6} &
					\text{4.363353048091063*${}^{\wedge}$7} \\
					\hline
					\text{Rank} & 6 & 7 & 8 \\
					\hline
					\text{Result} & \text{$1.761204119860\times 10^9$} & \text{$7.370817943138\times 10^{10}$} &
					\text{$3.1822242795255\times 10^{12}$} \\
					\hline
					\text{Rank} & 9 & 10 & 11 \\
					\hline
					\text{Result} & \text{$1.41158438406\times 10^{14}$} & \text{$6.412540767851\times 10^{15}$} &
					\text{$2.975347813775\times 10^{17}$} \\
					\hline
				\end{tabular}
			\end{table}

	The numerical Taylor expansion of the generating function is given as follows: 
	\begin{align}
		\textbf{GF}_{5\to5}(t)&=1.+28.14374534067907\cdot\mathcal{t}+891.8998182261959\cdot\mathcal{t}^2\notag\\
		&+30720.548700314932\cdot\mathcal{t}^3+1.1282643383555175\times 10^6\cdot\mathcal{t}^4\notag\\
		&+4.3633530480910465\times 10^7\cdot\mathcal{t}^5+1.7612041198600552\times 10^9\cdot\mathcal{t}^6\notag\\
		&+7.370817943138432\times 10^{10}\cdot\mathcal{t}^7+3.1822242795256416\times 10^{12}\cdot\mathcal{t}^8\notag\\
		&+1.411584384056077\times 10^{14}\cdot\mathcal{t}^9+6.412540767842304\times 10^{15}\cdot\mathcal{t}^{10}\notag\\
		&+2.975347813764606\times 10^{17}\cdot\mathcal{t}^{11}+\mathcal{O}(\mathcal{t}^{12}).
	\end{align}
	
		\item Pentagon to Box, $5\to4$
			
	We choose two typical integrals to verify our generating function. 
			\begin{table}[htbp]\centering\begin{tabular}{|c|c|c|c|}
				\hline
				\text{Rank} & 0 & 1 & 2 \\
				\hline
				\text{Result} & 0. & 1.9794571126964369 & 117.0165516634012 \\
				\hline
				\text{Rank} & 3 & 4 & 5 \\
				\hline
				\text{Result} & 5816.790907215448 & 278858.4717048013 & \text{$1.33561859165727\times 10^7$} \\
				\hline
				\text{Rank} & 6 & 7 & 8 \\
				\hline
				\text{Result} & \text{$6.467560554308\times 10^8$} & \text{$3.17978538594604\times 10^{10}$} &
				\text{$1.5892825779136\times 10^{12}$} \\
				\hline
				\text{Rank} & 9 & 10 & 11 \\
				\hline
				\text{Result} & \text{$8.07512917564463\times 10^{13}$} & \text{$4.16873605287026\times 10^{15}$} &
				\text{$2.18489096755686\times 10^17\cdot$} \\
				\hline
			\end{tabular}\caption{Reduction Result of Pentagon to $D_2D_3D_4D_5$ Triangle by \texttt{FIRE6}}\end{table}
		
			The numerical Taylor expansion of the generating function is given as follows: 
			\begin{align}
				\textbf{GF}&{}_{5\to4;\widehat{1}}(\mathcal{t})=
			1.979457112696436853508\cdot\mathcal{t}+     117.016551663401212089687\cdot\mathcal{t}^2 \notag\\
		&+5816.790907215448086546878\cdot\mathcal{t}^3+278858.471704801335516720233\cdot\mathcal{t}^4 \notag\\
	     &  +1.335618591657268161878\times 10^7\cdot\mathcal{t}^5 +6.467560554307838030352\times 10^8\cdot\mathcal{t}^6 \notag\\
		  & +3.179785385946040065817\times 10^{10}\cdot\mathcal{t}^7 +1.589282577913656415226\times 10^{12}\cdot\mathcal{t}^8 \notag\\
		  & +8.075129175644627972880\times 10^{13}\cdot\mathcal{t}^9 +4.168736052870256702551\times 10^{15}\cdot\mathcal{t}^{10}\notag\\
		  & +2.184890967556854672072\times 10^{17}\cdot\mathcal{t}^{11}+\mathcal{O}(\mathcal{t}^{12}).
			\end{align}
			
			\begin{table}[htbp]\centering\begin{tabular}{|c|c|c|c|}
				\hline
				\text{Rank} & 0 & 1 & 2 \\
				\hline
				\text{Result} & 0. & -4.281411630475701 & -344.28306461178465 \\
				\hline
				\text{Rank} & 3 & 4 & 5 \\
				\hline
				\text{Result} & -22151.543708665336 & \text{$-1.3240016518207\times 10^6$} &
				\text{$-7.657550600363\times 10^7$} \\
				\hline
				\text{Rank} & 6 & 7 & 8 \\
				\hline
				\text{Result} & \text{$-4.3528832488\times10^9$} & \text{$-2.449234588\times10^{11}$}
				& \text{$-1.36907010\times 10^{13}$} \\
				\hline
				\text{Rank} & 9 & 10 & 11 \\
				\hline
				\text{Result} & \text{$-7.61811824461\times10^{14}$} & \text{$-4.225067193912\times 10^{16}$}
				& \text{$-2.33742672788\times 10^{18}$} \\
				\hline
			\end{tabular}\caption{Reduction Result of Pentagon to $D_1D_2D_4D_5$ Triangle by \texttt{FIRE6}}\end{table}
		
			The numerical Taylor expansion of the generating function is given as follows: 
			\begin{align}
				\textbf{GF}{}_{5\to4;\widehat{3}}(\mathcal{t})&=
				 -4.281411630475701002407\cdot\mathcal{t}-344.283064611784617369902\cdot\mathcal{t}^2 \notag\\
			 &-22151.543708665336258844567\cdot\mathcal{t}^3 -1.324001651820724484404\cdot\mathcal{t}^4 \notag\\
			   &  -7.657550600362864047\times 10^{7}\cdot\mathcal{t}^5-4.35288324879994\times 10^{9}\cdot\mathcal{t}^6 \notag\\
			    & -2.4492345879991\times 10^{11}\cdot\mathcal{t}^7 -1.369070967619795958\times 10^{13}\cdot\mathcal{t}^8 \notag\\
			    & -7.61811824460816178\times 10^{14}\cdot\mathcal{t}^9 -4.2250671939123482\times 10^{16}\cdot\mathcal{t}^{10} \notag\\
			    & -2.337426727880305159\times 10^{18}\cdot\mathcal{t}^{11}+\mathcal{O}(\mathcal{t}^{19}).
			\end{align}

		\item Pentagon to Triangle, $5\to3$
			
	We choose two typical integrals to verify our generating function. 
			\begin{table}[htbp]\centering\begin{tabular}{|c|c|c|c|}
				\hline
				\text{Rank} & 0 & 1 & 2 \\
				\hline
				\text{Result} & 0. & 0. & 7.3203111013775555 \\
				\hline
				\text{Rank} & 3 & 4 & 5 \\
				\hline
				\text{Result} & 873.0823162422049 & 76037.34249486253 & \text{$5.893055817313\times 10^6$} \\
				\hline
				\text{Rank} & 6 & 7 & 8 \\
				\hline
				\text{Result} & \text{$4.32079052040\times 10^8$} & \text{$3.074305412771\times 10^{10}$} &
				\text{$2.14894585343427\times 10^{12}$} \\
				\hline
				\text{Rank} & 9 & 10 & 11 \\
				\hline
				\text{Result} & \text{$1.485324405683\times 10^{14}$} & \text{$1.0188953358844\times 10^{16}$}
				& \text{$6.9518264510604\times 10^{17}$} \\
				\hline
			\end{tabular}\caption{Reduction Result of Pentagon to $D_3D_4D_5$ Triangle by \texttt{FIRE6}}\end{table}
	
		The numerical Taylor expansion of the generating function is given as follows: 
			\begin{align}
			\textbf{GF}{}_{5\to3;\widehat{12}}(\mathcal{t})&=
			7.32031110138\cdot\mathcal{t}^2+873.082316242\cdot\mathcal{t}^3+76037.34249486\cdot\mathcal{t}^4\notag\\
			&+5.89305581731\times 10^6\cdot\mathcal{t}^5+4.3207905204\times 10^8\cdot\mathcal{t}^6\notag\\
			&+3.07430541277\times 10^{10}\cdot\mathcal{t}^7+2.14894585343\times 10^{12}\cdot\mathcal{t}^8\notag\\
			&+1.485324405683\times 10^{14}\cdot\mathcal{t}^9+1.018895335884\times 10^{16}\cdot\mathcal{t}^{10}\notag\\
			&+6.951826451060\times 10^{17}\cdot\mathcal{t}^{11}+\mathcal{O}(\mathcal{t}^{12}).
			\end{align}
			
			\begin{table}[htbp]\centering\begin{tabular}{|c|c|c|c|}
			\hline
			\text{Rank} & 0 & 1 & 2 \\
			\hline
			\text{Result} & 0. & 0. & -1.4748513263396505 \\
			\hline
			\text{Rank} & 3 & 4 & 5 \\
			\hline
			\text{Result} & -115.81178544031663 & -6395.139969561789 & -288631.67875521374 \\
			\hline
			\text{Rank} & 6 & 7 & 8 \\
			\hline
			\text{Result} & \text{$-1.0204216419115\times 10^7$} & \text{$-1.692522115271\times 10^8$}
			& \text{$1.550496634133\times 10^{10}$} \\
			\hline
			\text{Rank} & 9 & 10 & 11 \\
			\hline
			\text{Result} & \text{$2.369360641667\times 10^{12}$} & \text{$2.218100002242\times 10^{14}$} &
			\text{$1.770605970571\times 10^{16}$} \\
			\hline
		\end{tabular}\caption{Reduction Result of Pentagon to $D_1D_4D_5$ Triangle by \texttt{FIRE6}}\end{table}
	
		The numerical Taylor expansion of the generating function is given as follows: 
			\begin{align}
				\textbf{GF}&{}_{5\to3;\widehat{23}}(\mathcal{t})=
				-1.47485132634\cdot\mathcal{t}^2-115.81178544\cdot\mathcal{t}^3-6395.13996956\cdot\mathcal{t}^4\notag\\
				&-288631.6787552\cdot\mathcal{t}^5-1.02042164191\times 10^7\cdot\mathcal{t}^6-1.69252211527\times 10^8\cdot\mathcal{t}^7\notag\\
				&+1.550496634133\times 10^{10}\cdot\mathcal{t}^8+2.369360641667\times 10^{12}\cdot\mathcal{t}^9\notag\\
				&+2.2181000022\times 10^{14}\cdot \mathcal{t}^{10}+1.77060597057\times 10^{16}\cdot \mathcal{t}^{11}+\mathcal{O}(\mathcal{t}^{12}).
			\end{align}
	
		\item Pentagon to Bubble, $5\to2$
			
	We choose two typical integrals to verify our generating function. 
			\begin{table}[htbp]\centering\begin{tabular}{|c|c|c|c|}
			\hline
			\text{Rank} & 0 & 1 & 2 \\
			\hline
			\text{Result} & 0. & 0. & 0. \\
			\hline
			\text{Rank} & 3 & 4 & 5 \\
			\hline
			\text{Result} & -3.45043209014488 & -598.2368342483164 & -68719.25792386582 \\
			\hline
			\text{Rank} & 6 & 7 & 8 \\
			\hline
			\text{Result} & \text{$-6.5930722347316\times 10^6$} & \text{$-5.722281197191\times 10^8$} &
			\text{$-4.662849011137\times 10^{10}$} \\
			\hline
		\end{tabular}\caption{Reduction Result of Pentagon to $D_4D_5$ Bubble by \texttt{FIRE6}}\end{table}
	
		The numerical Taylor expansion of the generating function is given as follows: 
			\begin{align}
			\textbf{GF}&{}_{5\to2;\widehat{123}}(\mathcal{t})=-3.45043209014488\cdot\mathcal{t}^3-598.236834248316468\cdot\mathcal{t}^4\notag\\
			&-68719.2579238658296\cdot \mathcal{t}^5-6.593072234731597\times 10^6\cdot\mathcal{t}^6\notag\\
			&-5.722281197191451\times 10^8\cdot\mathcal{t}^7-4.662849011137064\times 10^{10}\cdot\mathcal{t}^8+\mathcal{O}(\mathcal{t}^9).
			\end{align}
			
			\begin{table}[htbp]\centering\begin{tabular}{|c|c|c|c|}
			\hline
			\text{Rank} & 0 & 1 & 2 \\
			\hline
			\text{Result} & 0. & 0. & 0. \\
			\hline
			\text{Rank} & 3 & 4 & 5 \\
			\hline
			\text{Result} & 13.319303811986032 & 948.0673270255978 & 58591.314814083744 \\
			\hline
			\text{Rank} & 6 & 7 & 8 \\
			\hline
			\text{Result} & \text{$3.3075884251460\times 10^6$} & \text{$1.7962165060926\times 10^8$} &
			\text{$9.51718360623\times 10^9$} \\
			\hline
		\end{tabular}\caption{Reduction Result of Pentagon to $D_1D_4$ Bubble by \texttt{FIRE6}}\end{table}

	The numerical Taylor expansion of the generating function is given as follows: 
			\begin{align}
			\textbf{GF}&{}_{5\to2;\widehat{235}}(\mathcal{t})=13.3193038119860332485\cdot\mathcal{t}^3+948.06732702559779935\cdot\mathcal{t}^4\notag\\
	&+58591.3148140837440432\cdot\mathcal{t}^5+3.307588425146047\times 10^6\cdot\mathcal{t}^6\notag\\
		&+1.796216506092603\times 10^8\cdot\mathcal{t}^7+9.5171836062295834\times 10^9\cdot\mathcal{t}^8+\mathcal{O}(\mathcal{t}^9).
			\end{align}
	
		\item Pentagon to Tadpole $5\to1$
	
	We choose two typical integrals to verify our generating function. 
			\begin{table}[htbp]\centering\begin{tabular}{|c|c|c|c|}
				\hline
				\text{Rank} & 0 & 1 & 2 \\
				\hline
				\text{Result} & 0. & 0. & 0. \\
				\hline
				\text{Rank} & 3 & 4 & 5 \\
				\hline
				\text{Result} & 0. & -61.3501671673956 & -7960.61229768613 \\
				\hline
				\text{Rank} & 6 & 7 & 8 \\
				\hline
				\text{Result} & -747095.8262283974 & \text{$-6.1530993935739\times 10^7$} &
				\text{$-4.746641454844\times 10^9$} \\
				\hline
			\end{tabular}\caption{Reduction Result of Pentagon to $D_4$ Tadpole by \texttt{FIRE6}}\end{table}
		
			The numerical Taylor expansion of the generating function is given as follows: 
			\begin{align}
	\textbf{GF}&{}_{5\to1;\widehat{1235}}(\mathcal{t})= -61.3501671673955991\cdot \mathcal{t}^4-7960.6122976861288296\cdot\mathcal{t}^5\notag\\
	  &-747095.8262283975179056\cdot\mathcal{t}^6-6.1530993935738919\times 10^7\cdot\mathcal{t}^7\notag\\&-4.7466414548441339\times 10^9\cdot\mathcal{t}^8 +\mathcal{O}(\mathcal{t}^9).
	\end{align}
					
			\begin{table}[htbp]\centering\begin{tabular}{|c|c|c|c|}
		\hline
		\text{Rank} & 0 & 1 & 2 \\
		\hline
		\text{Result} & 0. & 0. & 0. \\
		\hline
		\text{Rank} & 3 & 4 & 5 \\
		\hline
		\text{Result} & 0. & -0.4647775732011648 & -76.65585039332824 \\
		\hline
		\text{Rank} & 6 & 7 & 8 \\
		\hline
		\text{Result} & -7394.867372623504 & -504749.27267807984 & \text{$-2.9542488797636\times 10^7$} \\
		\hline
	\end{tabular}\caption{Reduction Result of Pentagon to $D_1$ Tadpole by \texttt{FIRE6}}\end{table}
  		
  			 The numerical Taylor expansion of the generating function is given as follows: 
     		\begin{align}
     	\textbf{GF}{}_{5\to1;\widehat{2345}}(\mathcal{t})&=-0.46477757320116485\cdot\mathcal{t}^4-76.65585039332823\cdot\mathcal{t}^5\notag\\&-7394.8673726235035272\cdot\mathcal{t}^6-504749.27267807984127\cdot\mathcal{t}^7\notag\\&-2.95424887976361\times 10^7\cdot\mathcal{t}^8+\mathcal{O}(\mathcal{t}^9).
     \end{align}
	\end{itemize}
\end{appendices}

	\appendix
	

	\bibliographystyle{JHEP}
    \bibliography{reference}

\end{document}